\documentclass[3p,12pt]{elsarticle}
\usepackage{threeparttable,booktabs,tabularx}
\usepackage[fleqn]{amsmath}
\usepackage{cases}
\usepackage{multirow}
\usepackage{xcolor}
\usepackage[normalem]{ulem}
\usepackage{tabularx}
\usepackage{siunitx}
\usepackage[caption=false]{subfig}
\usepackage{comment}
\usepackage{algorithm}
\usepackage{algpseudocode}
\usepackage{algorithmicx}
\usepackage{grffile}
\usepackage{rotating}
\usepackage{array}
\usepackage{lineno}
\usepackage{hyperref}
\usepackage{makecell}
\usepackage{graphicx,enumerate}
\usepackage{amssymb}
\usepackage{bm,amsmath}
\usepackage[caption=false]{subfig}
\usepackage{caption,color}
\usepackage{threeparttable}
\usepackage{subeqnarray}
\usepackage{multirow}
\usepackage{lscape}
\usepackage{rotating}
\usepackage{graphics}
\floatname{algorithm}{Algorithm}
\usepackage{epstopdf}
\journal{Journal of Computational Physics}

\biboptions{numbers,sort&compress} 

\newcommand{\Kn}{\text{Kn}}

\begin{document}
\begin{frontmatter}
\title{Boosting the convergence of DSMC by GSIS}
\author[1]{Liyan Luo}
\author[1]{Qi Li} 
\ead{liq@sustech.edu.cn}
\author[2]{Fei Fei} 
\ead{ffei@hust.edu.cn}
\author[1]{Lei Wu}
\address[1]{Department of Mechanics and Aerospace Engineering,
Southern University of Science and Technology, 518055 Shenzhen, China}
\address[2]{School of Aerospace Engineering, Huazhong University of Science and Technology, 430074 Wuhan, China}

\begin{abstract}
A deterministic-stochastic coupling scheme is developed for simulating rarefied gas flows, where the key process is the alternative solving of the macroscopic synthetic equations [Su \textit{et al.}, J.~Comput.~Phys., 407 (2020) 109245] and the mesoscopic equation via the asymptotic-preserving time-relaxed Monte Carlo scheme [Fei, J.~Comput.~Phys., 486 (2023) 112128]. Firstly, the macroscopic synthetic equations are exactly derived from the Boltzmann equation, incorporating not only the Newtonian viscosity and Fourier thermal conduction laws but also higher-order constitutive relations that capture rarefaction effects; the latter are extracted from the stochastic solver over a defined sampling interval. Secondly, the macroscopic synthetic equations, with the initial field extracted from the stochastic solver over the same sampling interval, are solved to the steady state or over a certain iteration steps. Finally, the simulation particles in the stochastic solver are updated to match the density, velocity, and temperature obtained from the macroscopic synthetic equations. Moreover, simulation particles in the subsequent interval will be partly sampled according to the solutions of macroscopic synthetic equations. As a result, our coupling strategy enhances the asymptotic-preserving characteristic of the stochastic solver and substantially accelerates convergence towards the steady state. Several numerical tests are performed, and it is found that our method can reduce the computational cost in the near-continuum flow regime by two orders of magnitude compared to the direct simulation Monte Carlo method.
\end{abstract}


\end{frontmatter}
\section{Introduction}\label{sec:1}

Rarefied gas flows are common in contemporary engineering fields, ranging from the aerodynamics of space re-entry capsules~\cite{votta-2013} to the microelectromechanical systems~\cite{titov-2008}. These flows are characterized by the Knudsen number Kn, which is the ratio of the molecular mean free path $\lambda$ to the characteristic system length $L$.  
In the continuum flow regime ($\text{Kn}<0.001$), the gas flow can be modeled using the Navier-Stokes (NS) equations. 
However, as the Knudsen number increases, gas molecules exhibit significant departures from thermodynamic equilibrium. Consequently, the NS solutions begin to deviate substantially from experimental observations due to the limitations of Newtonian viscosity and Fourier thermal conduction laws \cite{alsmeyer-1976}.
To accurately account for rarefaction effects, it is necessary to employ the Boltzmann equation. In fact, according to the Chapman-Enskog expansion, the NS equations are a specific solution of the Boltzmann equation applicable in the continuum flow limit \cite{chapman-1990, struchtrup-2006}. Solving the Boltzmann equation is considered the most direct method for multiscale problems, as it fundamentally encompasses the continuum, slip, transition, and free-molecular flow regimes.

Two numerical methods are prevailing in the field of rarefied gas dynamics: the deterministic discrete velocity method \cite{aristov-2001} and the stochastic direct simulation Monte Carlo (DSMC) method \cite{bird-1994}. In deterministic approaches, the velocity distribution function (VDF) in the Boltzmann equation is discretized in both spatial and velocity dimensions, allowing for direct numerical solution by the sophisticated computational fluid dynamics methods. However, in scenarios with large Knudsen numbers or in hypersonic flows, the number of velocity grids required escalates significantly, leading to substantial demands on both computer memory and processing time. 
In contrast, the DSMC employs simulation particles to represent real gas molecules, and there are only dozens particles in one spatial cell, each representing a huge number of real molecules. The DSMC is efficient in simulating rarefied hypersonic flows, but it is susceptible to statistical fluctuations, making it less suitable for low-speed and/or time-dependent flows. Moreover, as the Knudsen number diminishes, both methods become computationally prohibitive. For instance, to reduce the numerical dissipation, the grid size should be about one third of the molecular mean free path in DSMC, leading to an increased number of cells in the near-continuum regime. The situation is further exacerbated by the necessity for the time step to be smaller than the mean collision time, where a substantial number of time evolution is needed before sampling the steady-state solution.

Many strategies have been proposed to improve the efficiency of DSMC in the near-continuum flow regime.
To mitigate the computational load, a hybrid NS-DSMC algorithm has been developed \cite{schwartzentruber-2006,schwartzentruber-2007}, where the computational domain is divided into a continuum flow region and a rarefied one, enabling the application of NS equations and DSMC method, respectively. This hybrid approach has been shown to enhance the computational efficiency for multi-scale problems. However, in many engineering applications, delineating the boundary between the two regions can be challenging. 
To avoid the domain decomposition, the time-relaxed Monte Carlo (TRMC) method~\cite{pareschi-2001,Gabetta-1997}, the exponential Runge-Kutta~\cite{dimarco-2011} and the asymptotic-preserving (AP) Monte Carlo method~\cite{ren-2014} have been developed.  
These methods enable a larger time step, recover the kinetic scheme of Boltzmann equation when the Knudsen number is large, while approach the local Maxwell distribution when Kn$\rightarrow 0$. However, they only preserve the Euler asymptotics~\cite{jin-1999}.

Recently, the AAP-TRMC that accurately preserves the NS equations has been established~\cite{fei-2023}, where the Boltzmann collision operator is also simulated by the Wild sum expansion~\cite{wild-1951}, but the micro-macro decomposition of the collision operator is based on the Chapman-Enskog expansion. That is, when Kn is small, the macroscopic collision part is calculated by a Crank-Nicolson scheme to achieve second-order accuracy and the NS asymptotics~\cite{fei-2020}, and when Kn is large, it becomes the original TRMC method. 
Several numerical tests are performed to show that the AAP-TRMC has lower numerical dissipation and higher computational efficiency for multi-scale flow simulations. 


Given that the majority of issues in rarefied gas dynamics pertain to steady-state solutions, the tracking of time evolution is not a requisite. Recently, the deterministic general synthetic iterative scheme (GSIS) has been introduced, offering an accurate method for solving the Boltzmann equation~\cite{su-2020-can,zeng-2023-cicp}. This approach circumvents the laborious time evolution process and significantly reduces computational time by several orders of magnitude when the Knudsen number is low. The core concept of GSIS is to iteratively solve both the mesoscopic Boltzmann equation and the macroscopic synthetic equations that are precisely derived from the Boltzmann equation. The Boltzmann equation supplies the synthetic equations with higher-order constitutive relations, while the synthetic equations, which can be rapidly solved to the steady state, direct the evolution of the VDF within the Boltzmann equation. Through this two-way coupling, fast convergence and AP are achieved~\cite{su-2020}. 

Recently, the GSIS has been successfully extended to enhance the convergence rate of the low-variance DSMC method~\cite{luo-2023}. For instance, at a Knudsen number of 0.01, the deterministic-stochastic coupling approach requires only 100 spatial cells and $10^4$  evolution steps to obtain the steady-state solution for linearized Poiseuille flow, as opposed to the 300 cells and $10^5$ steps required by the low-variance DSMC. 
The objective of this research is to extend GSIS to expedite the convergence of the DSMC method for nonlinear rarefied gas flows. Such an extension would be non-trivial, as the linearized Bhatnagar-Gross-Krook equation is solved in the low-variance DSMC, where the AP property is preserved through a direct modification on the gain term in the simplified collision operator. In contrast, the DSMC method solves the Boltzmann equation where the collision operator is complicated, leading to a significant challenge for the reciprocal feedback with the synthetic equations.




The remaining part of this paper is organized as follows. Different stochastic methods for solving the Boltzmann equation are reviewed in Section~\ref{sec:2}; the GSIS and its strategy of boosting convergence are stated in Section~\ref{sec:3}; the numerical scheme for solving the macroscopic synthetic equations is presented in Section~\ref{sec:4}; the efficiency and accuracy of the proposed deterministic-stochastic coupling algorithm is assessed based on several standard multiscale flow problems in Section~\ref{sec:5}. Finally, the conclusion as well as future outlooks are presented in Section~\ref{sec:6}.
 \section{Stochastic particle method for solving Boltzmann equation}\label{sec:2}

In this section, three stochastic particle methods for solving the Boltzmann equation are reviewed, i.e., the DSMC, TRMC, and AAP-TRMC methods.

\subsection{Boltzmann equation}

The Boltzmann equation can be written in the dimensionless form as follows \cite{bird-1994}: 
\begin{equation}
\frac{\partial f(t,\bm{x},\bm{v})}{\partial t}+\bm{v}\cdot\frac{\partial f(t,\bm{x},\bm{v})}{\partial\bm{x}}=\frac{1}{\epsilon}Q(f,f),
\label{eq:boltzmann_equation}
\end{equation}
where $f(t,\bm{x},\bm{v})$ is the velocity distribution function at position $\bm{x}$ with molecular velocity $\bm{v}$ and time $t$. 
All variables are written in their dimensionless forms related to reference length $L_0$, reference density $\rho_0$, reference temperature $T_0$, and most probable speed $c_0=\sqrt{k_BT_0/m}$, where $k_B$ and $m$ represent the Boltzmann constant and molecular mass, respectively.
Important parameter $\epsilon$ in the right-hand side of Eq.~(\ref{eq:boltzmann_equation}) is known as the Knudsen number, and the binary collision operator $Q(f,f)$ can be described as:
\begin{equation}
\begin{aligned}
Q(f,f) = \underbrace{\int_{\mathbb{R}^3}\int_{\mathbb{S}^2}f(\bm{v}')f(\bm{v}_*')B(c_r,\omega)d\omega d\bm{v}_* }_{Q^+(f,f)}-
\underbrace{\int_{\mathbb{R}^3}\int_{\mathbb{S}^2}f(\bm{v})f(\bm{v}_*)B(c_r,\omega) d\omega d\bm{v}_*}_{Q^-(f)},
\end{aligned}
\label{eq:collision_operater}
\end{equation}
where $\bm{v}$, $\bm{v}_*$ represent the pre-collisional velocities of a collision pair, and $\bm{v}'$, $\bm{v}_*'$ are their post-collisional velocities. $c_r=|\bm{v}-\bm{v}_*|$ is the relative velocity of the collision pair and $\omega$ is the solid angle. $Q^+$ and $Q^-$ denote the gain term and loss term of the collision operator, respectively.
The non-negative kernel function $B(c_r,\omega)$ is determined by the intermolecular potential. In the following paper, the variable hard sphere (VHS) model is employed for simplicity~\cite{bird-1994}.




The Boltzmann collision operator conserves the mass, momentum and energy:
\begin{equation}
\int_{\mathbb{R}^3}Q(f,f)\phi (\bm{v})d\bm{v} = 0, \,\, \text{for }\phi (\bm{v})=1,\bm{v},|\bm{v}|^2.
\label{eq:Qconserve}
\end{equation}
In addition, it satisfies a well-known $H$-theorem $
\int_{\mathbb{R}^3}Q(f,f)\ln(f)d\bm{v}\leq 0$,
which implies that the non-trival distribution function $f$ for $Q(f,f)=0$ has the local Maxwellian distribution, i.e.,
\begin{equation}
f_M(\rho,\bm{u},T) = \frac{\rho}{\sqrt{2\pi T}^3}\exp\left(-\frac{\bm{c}^2}{2T}\right),
\label{eq:Maxwellian}
\end{equation}
where the dimensionless variables $\rho$, $\bm{u}$, $T$ are the density, mean velocity and temperature of the gas field and $\bm{c}=\bm{v}-\bm{u}$ is the peculiar velocity. Moreover, the macroscopic properties, such as the stress tensor $\sigma_{ij}$ (normalized by $\rho_0(k_B/m)T_0$) and heat flux $\bm{q}$ (normalized by $\rho_0(k_B/m)T_0c_0$), can be obtained by taking the moments of the VDF:
\begin{equation}
\begin{aligned}
&\rho = \int_{\mathbb{R}^3}fd\bm{v},\qquad\bm{u} = \frac{1}{\rho}\int_{\mathbb{R}^3}\bm{v}fd\bm{v},\qquad T = \frac{1}{3\rho}\int_{\mathbb{R}^3}\bm{c}^2fd\bm{v}, \\
&\sigma_{ij}=\int_{\mathbb{R}^3}c_{\langle i}c_{j\rangle}fd\bm{v}, \qquad
\bm{q}=\frac{1}{2}\int_{\mathbb{R}^3}\bm{c}\bm{c}^2fd\bm{v}.
\end{aligned}
\label{eq:rhout}
\end{equation}
where the angle brackets $\langle i,\,j\rangle$ stands for the trace-less component of a tensor.

\subsection{The DSMC method}

DSMC decouples the particle motion and collision, therefore, the Boltzmann equation is numerically split into the advection and collision steps:
\begin{equation}
\begin{aligned}
&\text{Advection:}\qquad \frac{\partial f}{\partial t}+\bm{v}\cdot\frac{\partial f}{\partial \bm{x}}=0.\\
&\text{Collision:}\qquad \left[\frac{\partial f}{\partial t}\right]_{\text{coll}}=\frac{1}{\epsilon}Q(f,f).
\label{eq:splitting_eq}
\end{aligned}
\end{equation}
During the advection step, the particle velocities remain unchanged, but their positions is updated in accordance with their velocities. Thus, the solution of the advection step is written as:
\begin{equation}
f^*(\bm{v};\bm{x})=f^n(\bm{v};\bm{x}-\bm{v}\Delta t),
\end{equation}
where $f^*(\bm{v};\bm{x})$ represents the distribution function after the advection process and $f^n(\bm{v};\bm{x})$ denotes the distribution function at time $n\Delta t$. 

After that, the collision occurs in each cell, where particle pairs collide based on certain probabilities without changing positions. The most efficient sampling algorithms for collisions in DSMC are Bird's no time counter scheme and the Nanbu-Babovsky scheme~\cite{nanbu-1980,babovsky-1986}. Specifically, using the Nanbu-Babovsky scheme, the Boltzmann equation can be rewritten as:
\begin{equation}
\left[\frac{\partial f}{\partial t}\right]_\text{coll}=\frac{1}{\epsilon}\left[P(f,f)-\beta f\right].
\label{eq:collisionseperate}
\end{equation}
For the VHS model, the non-negative bilinear operator $P(f,f)$ based on the upper bound of kernel $\bar{B}=\mathop{\max}_{ij} B(|\bm{v}_i-\bm{v}_j|)$ (where $i,\,j$ represent different collision particles) can be constructed as follows:
\begin{equation}
\begin{aligned}
&P(f,f)=Q^+(f,f)+f(\bm{v})\int_{\mathbb{R}^3}\int_{\mathbb{S}^2}\left[\bar{B}-B(c_r,\omega)\right]f(\bm{v}_*)d\omega d\bm{v}_*, \\ 
&\beta =\int_{\mathbb{R}^3}\int_{\mathbb{S}^2} \bar{B}f(\bm{v}_*)d\omega d\bm{v}_*= \rho \bar{\sigma}\quad\text{with} \quad \bar{\sigma}=\int_{\mathbb{S}^2}{\bar{B}d\omega}=4\pi \bar{B}.
\end{aligned}
\end{equation}
In Eq.~\eqref{eq:collisionseperate}, $\beta$ is an upper bound of the coefficient of the loss term $Q^-$, which satisfies the condition that $\beta f\ge Q^-(f,f)$. 
The choice of the upper bound $\beta$ should ideally be minimized to augment computational efficiency during the collision process. Since the maximum relative velocity between collision pairs is different in each cells, the value of $\beta$ varies from cell to cell during the numerical simulation. 

By applying the forward Euler scheme, the time discretization of Eq.~\eqref{eq:collisionseperate} can be written as:
\begin{equation}
f^{n+1}=\underbrace{\left(1-\frac{\beta\Delta t}{\epsilon}\right)f^*+\frac{\beta\Delta t}{\epsilon}\frac{P(f^*,f^*)}{\beta}}_{\mathcal{C}^{DSMC}_{\Delta t}\left(f^*\right)},
\label{eq:timediscretize}
\end{equation}
where $\mathcal{C}^{DSMC}_{\Delta t}$ denotes the collision operator of DSMC within the time step $\Delta t$. According to Eq.~\eqref{eq:timediscretize}, during the collision step, particles collide with the probability $\beta\Delta t/\epsilon$. 
Note that this probabilistic interpretation fails if $\Delta t/\epsilon$ is so large that the first term in the right hand side of Eq.~\eqref{eq:timediscretize} becomes negative.
Therefore, in the near-continuum regime, the traditional DSMC method is inadequate in producing results effectively and accurately. To overcome this defect, several DSMC schemes with the AP property have been developed. The AP property allows the scheme to employ larger time step and cell size than DSMC, and recover the Euler solution when $\epsilon\rightarrow0$.

\subsection{Time-relaxed Monte Carlo method}

One of the DSMC schemes with AP property is the time-relaxed Monte Carlo method proposed by Pareschi~\cite{pareschi-2001}. By using the transformation:
\begin{equation}
\tau=(1-e^{-\beta t/\epsilon}),\qquad F(v,\tau)=f(\bm{v},t)e^{\beta t/\epsilon},
\end{equation}
the collision step for Eq.~\eqref{eq:collisionseperate} can be rewritten as:
\begin{equation}
\frac{\partial F}{\partial \tau}=\frac{1}{\beta}P(F,F).
\end{equation}
According to the Wild sum \cite{wild-1951}, the numerical solution during collision process can be expressed by:
\begin{equation}
f^{n+1}=\underbrace{\sum_{k=0}^mA_kf_k+A_{m+1}f_M^n}_{\mathcal{C}^{TRMC}_{\Delta t}},
\label{eq:standardtrmc}
\end{equation}
where $\mathcal{C}^{TRMC}_{\Delta t}$ denotes the collision operator of TRMC. In Eq.~\eqref{eq:standardtrmc}, the parameter $m$ denotes the $m$-th truncation ($m\ge 1$). And the functions $f_k$ are given according to the recurrence formula:
\begin{equation}
f_{k+1}(\bm{v})=\frac{1}{k+1}\sum^k_{h=0}\frac{1}{\beta}P(f_h,f_{k-h}), \qquad k=0,1,\cdots,
\end{equation}
and $f_0(\bm{v})=f^*(\bm{v};\bm{x})$. The coefficients $A_k$ are non-negative weight functions:
\begin{equation}
\begin{aligned}
&A_k=e^{-\beta\Delta t/\epsilon}(1-e^{-\beta\Delta t/\epsilon})^k,\qquad k=0,\cdots,m-1, \\
&A_m=1-\sum_{k=0}^{m-1} A_k-A_{m+1}, \qquad A_{m+1}=(1-e^{-\beta\Delta t/\epsilon})^{m+2}.
\end{aligned}
\label{eq:weightfunction}
\end{equation}
Note that in Eq.~\eqref{eq:standardtrmc}, the summation of coefficient functions $A_k$ is equal to 1. This signifies the conservation of particles, where the total probability for different collision processes involving different particles sums to 1. And a certain part of full collision process is replaced by sampling from Maxwellian, which reduces the computational time in the collision process compared to traditional DSMC method. In addition, as shown in Eq.~\eqref{eq:weightfunction}, if the time step is fixed, when the Knudsen number approaches zero ($\epsilon\rightarrow 0$), the probability of particles sampling from local Maxwellian is nearly 1 ($A_{m+1}\rightarrow 1$). This implies that in the near-continuum regime, the vast majority of particles are sampled from the local Maxwellian, and the distribution function $f^{n+1}$ relaxes immediately to its equilibrium distribution. Thus, the TRMC scheme only preserves the Euler asymptotic, where the shear stress and the heat flux can not be accurately captured in the fluid limit ($\beta\Delta t/\epsilon \gg 1$). 

\subsection{TRMC scheme preserving Navior-Stokes asymptotics}

As discussed in previous section, the standard TRMC scheme satisfies AP property to the Euler solution and only has the first order of accuracy as $\epsilon \rightarrow 0$. 
To achieve an AP property for NS solution, the AAP-TRMC method has been developed. Using the Chapman-Enskog expansion~\cite{chapman-1990}, the AAP-TRMC method divides the collision operator into macro and micro parts. The macro part is reconstructed from the first order of Chapman-Enskog  expansion and solved using a second order scheme. Meanwhile, the micro part which refers to the higher order of the Chapman-Enskog  expansion remains unchanged and solved by the standard TRMC scheme.
The numerical scheme of the collision operator for AAP-TRMC scheme is written as \cite{fei-2023}:
\begin{equation}\label{eq:AAPTRMC_collsion_operator}
\begin{aligned}
f^{n+1}=\underbrace{\sum_{k=0}^mA_kf_k+A_{m+1}\left(f_M^{n+1/2}+\theta f_\alpha\right)}_{\mathcal{C}^{AAP-TRMC}_{\Delta t}}, 
\end{aligned}
\end{equation}
where $\mathcal{C}^{AAP-TRMC}_{\Delta t}$ denotes the collision operator of the AAP-TRMC scheme and $\theta$ is a scaling factor to maintain the positivity of the VDF $\left(f_M^{n+1/2}+ f_\alpha\right)$:
\begin{equation}
\theta = \begin{cases}1,\qquad &\beta\Delta t/\epsilon \ge 1, \\ \min \left\{1,\frac{\epsilon}{\beta\Delta t}\right\}, \qquad &\beta\Delta t/\epsilon < 1, \end{cases}  
\end{equation}
In Eq.~\eqref{eq:AAPTRMC_collsion_operator}, the distribution function $f_\alpha$ is given by
\begin{equation}
\begin{aligned}
f_\alpha =\frac{f_M^{n+1/2}}{A_{m+1}}&\left\{\left[1-\frac{\Delta t}{2\mu/p}-\left(1+\frac{\Delta t}{2\mu/p}\right)\Phi_1\right]\frac{c_{\langle i}c_{j\rangle}\sigma_{ij,NS}}{2pT}+\right. \\&\left.\left[1-\frac{\text{Pr}\Delta t}{2\mu/p}-\left(1+\frac{\text{Pr}\Delta t}{2\mu/p}\right)\Phi_2\right] \frac{2c_kq_{k,NS}}{5pT}\left(\frac{c^2}{2T}-\frac{5}{2}\right)\right\},
\end{aligned}
\label{eq:f_alpha}
\end{equation}
where the parameter $\Phi_1$ and $\Phi_2$ are coefficient functions associated with the time step as well as the order of the AAP-TRMC scheme. Specifically, for the second order AAP-TRMC scheme (AAP-TRMC-2nd), the forms of these coefficient functions are given by
\begin{equation}
\begin{aligned}
\Phi_1 &= A_0 + \left(A_1 + \frac{1}{2}A_2\right)\left(1-\frac{1}{\chi}\right)+\frac{1}{2}A_2\left(1-\frac{1}{\chi}\right)^2, \\
\Phi_2 &= A_0 + \left(A_1 + \frac{1}{2}A_2\right)\left(1-\frac{\text{Pr}}{\chi}\right)+\frac{1}{2}A_2\left(1-\frac{\text{Pr}}{\chi}\right)^2,
\end{aligned}
\label{eq:coefficientfunctions_AAPTRMC}
\end{equation}
where the parameter $\chi = \mu\beta/(p\epsilon)$. In this study, we use the AAP-TRMC-2nd scheme for sake of simplicity. Details can be found in Ref.~\cite{fei-2023}.


According to Eq.~\eqref{eq:AAPTRMC_collsion_operator}, except for the sampling distribution function, $\theta f_\alpha$, other collision processes in AAP-TRMC are the same as the original TRMC scheme. This implies that in the kinetic limit ($A_{m+1}\rightarrow 0$), the AAP-TRMC scheme reduces to TRMC scheme. Additionally, in the fluid limit ($\epsilon \rightarrow 0$), particles with a probability of $A_{m+1}$ within a single cell are sampled based on the VDF $f_M^{n+1/2}+\theta f_\alpha$, rather than the Maxwellian distribution function in TRMC, see Eq.~\eqref{eq:f_alpha}. Therefore, the AAP-TRMC asymptotically preserves the NS limit. By applying the acceptance-rejection algorithm~\cite{garcia-1998} or the Metropolis-Hastings method~\cite{pfeiffer-2018}, particle velocities can be easily sampled to satisfy the VDF above.

 \section{The particle-based general synthetic iterative scheme}\label{sec:3}
In this section, the GSIS is coupled with AAP-TRMC to further enhance the computational efficiency of DSMC in the near-continuum regime. The macroscopic properties are updated in GSIS and employed as the continuous guidance for the particle evolution. Two ways of guidance are introduced in this section. Firstly, the particle information undergoes a direct modification through a linear transformation according to the updated macroscopic properties. Secondly, the sampling distribution function in the collision of AAP-TRMC is tailored to align with the updated macroscopic properties during the subsequent sampling interval.

\subsection{Macroscopic synthetic equations}

By multiplying Eq.~\eqref{eq:boltzmann_equation} with $\phi$ in Eq~\eqref{eq:Qconserve} and integrating them with respect to $d\bm{v}$, the macroscopic equations can be derived as:
\begin{equation}
\begin{aligned}
\frac{\partial \rho}{\partial t}+\nabla\cdot(\rho \bm{u})=0&, \\
\frac{\partial \rho\bm{u}}{\partial t}+\nabla\cdot(\rho\bm{u}\bm{u})+\nabla p+\nabla\cdot\bm{\sigma}=0&,\\
\frac{\partial \rho E}{\partial t}+\nabla\cdot\left(\rho E\bm{u}+p\bm{u}+\bm{u}\cdot\bm{\sigma}+\bm{q}\right)=0&,
\end{aligned}
\label{eq:Navior-Stokes}
\end{equation}
where $p=\rho T$ and  $E=\frac{3}{2}\rho T+\frac{1}{2}\rho u^2$.

The above equations are not closed becuase  the shear stress $\bm{\sigma}$ and the heat flux $\bm{q}$ cannot be expressed in terms of low order moments. According to the Chapman-Enskog expansion, the linear constitutive relations for $\bm{\sigma}$ and $\bm{q}$ can be derived from the first-order expansion:
\begin{equation}
\begin{aligned}
&\sigma_{ij,\text{NS}} = -\mu\left(\frac{\partial u_i}{\partial x_j}+\frac{\partial u_j}{\partial x_i}-\frac{2}{3}\delta_{ij}\nabla\cdot\bm{u}\right),\\
&\bm{q}_{\text{NS}}=-\kappa \nabla T,
\end{aligned}
\label{eq:linear-constitutive-relations}
\end{equation}
where the heat conductivity $\kappa$ is determined by the viscosity $\mu$ and the Prandtl number (Pr): $\kappa = \mu c_p/\text{Pr}$. Generally, $c_p$ represents the heat capacity at the constant pressure giving that $c_p=5R/2$ in the monoatomic gas.


However, the NS constitutive relations of $\bm{\sigma}$ and $\bm{q}$ are not adequate for solving the macroscopic equations under the rarefied conditions, where the shear stress and the heat flux should be calculated directly from the velocity distribution function without any truncation.  In GSIS, $\bm{\sigma}$ and $\bm{q}$ are decomposed into two parts: the NS linear constitutive relations and the high-order terms obtained from the velocity distribution function explicitly, given by:
\begin{equation}
\begin{aligned}
\sigma_{ij}&=\sigma_{ij,\text{NS}}+\text{HoT}_{\sigma_{ij}},\\
\bm{q}&=\bm{q}_{\text{NS}}+\text{HoT}_{\bm{q}},
\end{aligned}
\label{eq:full_stress_heatflux}
\end{equation}
where the HoTs are directly constructed according to the basic definitions of the shear stress and heat flux~\cite{zhu-2021}:
\begin{equation}
\begin{aligned}
&\text{HoT}_{\sigma_{ij}}=\int f^*c_{\langle i}^*c_{j\rangle}^*d\bm{v}-\sigma_{ij,NS}^*,\\
&\text{HoT}_{\bm{q}}=\frac{1}{2}\int f^*\bm{c}^*\left(c^*\right)^2d\bm{v}-\bm{q}_{\text{NS}}^*,
\end{aligned}
\label{eq:highorderterms}
\end{equation}
and the superscript * represents the value obtained from the solution of the Boltzmann equation (or solver). 


In the stochastic particle method, the time-averaged macroscopic properties in a single cell are given by: 
\begin{equation}
\begin{aligned}
&\rho=\frac{N_{eff}}{V_{cell} }\frac{1}{N_s}\sum_{s=1}^{N_s}N_{p}^{(s)},\,\, u_i=\frac{1}{\sum\limits_{s=1}^{N_s}N_{p}^{(s)}}\sum_{s=1}^{N_s}\sum_{p=1}^{N_{p}^{(s)}}{v_{i,p}^{(s)}},\,\, T=\frac{1}{3\sum\limits_{s=1}^{N_s}N_{p}^{(s)}}\sum_{s=1}^{N_s}\sum_{p=1}^{N_{p}^{(s)}}\left|\bm{v}_p^{(s)}-\bm{u}\right|^2,\\
&\sigma_{ij}= \frac{\rho}{\sum\limits_{s=1}^{N_s}N_{p}^{(s)}}\sum_{s=1}^{N_s}\sum_{p=1}^{N_p^{(s)}}\left[\left(v_{i,p}^{(s)}-u_i\right)\left(v_{j,p}^{(s)}-u_j\right)-\frac{\delta_{ij}}{3}\left|\bm{v}_p^{(s)}-\bm{u}\right|^2\right],\\
&q_i=\frac{\rho}{2\sum\limits_{s=1}^{N_s}N_{p}^{(s)}}\sum_{s=1}^{N_s}\sum_{p=1}^{N_p^{(s)}}\left(v_{i,p}^{(s)}-u_i\right)\left|\bm{v}_p^{(s)}-\bm{u}\right|^2,
\end{aligned}
\label{eq:statisticmacro}
\end{equation}
where the parameter $N_{eff}$ is the number of real gas molecules represented by one simulated particle, and $N_p^{(s)}$ denotes the number of simulated particles $N_p$ within this cell for the $s$-th sample. Moreover, the parameter $N_s$ represents the time-averaged sampling interval. 
The selection of $N_s$ is associated with the time step $\Delta t$. 
When the time step is small (much smaller than the mean collision time), a relatively large value of $N_s$ should be chosen, producing smoother macroscopic properties for solving the synthetic equations. 
On the contrary, if the time step exceeds the mean collision time, which frequently occurs in the near-continuum regime, $N_s$ can be relatively small. 
In general, this averaging method helps reduce the fluctuations of macroscopic properties, thereby enhancing the stability of the subsequent macroscopic solver.

\subsection{Feedback of macroscopic properties to DSMC }

When the macroscopic synthetic equations are solved, the resulting solutions represent a closer approximation to the final steady state. Such information should be fed back to the kinetic solver. When applying the GSIS within the deterministic framework~\cite{su-2020-can,liu-2024}, the updated macroscopic properties $\left(\bm{M}^{n+1/2}=\left[\rho^{n+1/2},\bm{u}^{n+1/2},T^{n+1/2}\right]\right)$ explicitly correct the VDF $f^n$, and the corrected VDF $f^{n+1/2}$ continues to be updated iteratively in DVM. However, due to the nature of the stochastic particle method, there is no explicit expression of VDF. As a result, modifications to the VDF are directly manifested through alterations of particle information.


\subsubsection{Linear transformation on particle information}\label{sec:PVS}


For the sake of simplicity, the updated macroscopic properties are denoted as $\bm{M}^{**}$. First, the number of simulation particles within a single cell should be adjusted according to the updated number density. 
To this end, a simple replicating and discarding procedure are applied~\cite{degond-2011}. 
According to the updated number density $\rho^{**}$, the predicted number of particles in specific cell $N_p^{**}$, can be determined.
Since $N_p^{**}$ should be an exact integer, a suitable stochastic rounding technique is essential, given by: 
\begin{equation}
N_p^{**}=\text{Iround}\left(\frac{\rho^{**}V_{cell}}{N_{eff}}\right),
\label{eq:Npstarstar}
\end{equation}
where the stochastic rounding function $\text{Iround}(x)$ is defined as:
\begin{equation}
\text{Iround}(x)=\begin{cases} \left\lfloor x \right\rfloor+1, & \text{with probability } x-\left\lfloor x \right\rfloor, \\ \left\lfloor x \right\rfloor, &  \text{with probability } 1-x+\left\lfloor x \right\rfloor ,\end{cases}
\label{eq:Iround}
\end{equation}
where $\left\lfloor x \right\rfloor$ represents the integer part of $x$.
When the current number of particles in a specific cell $N_p$ is smaller than $N_p^{**}$ ($N_p < N_p^{**}$), $\Delta N_p = N_p^{**}-N_p$ simulated particles need to be replicated. The replicated particle velocities are assigned based on a random selection of particles within the current cell, and their positions are uniformly distributed across the cell. 
Conversely, when $N_p > N_p^{**}$, the discarding process will be executed. A number of $\Delta N_p=N_p-N_p^{**}$ simulated particles are randomly selected and subsequently eliminated. In general, the replicating and discarding process guarantee the number of simulation particles within a single cell to satisfy the requirement of updated density. 

Second, the temporary velocity $\bm{u}^{(t)}$ and temperature $T^{(t)}$ are calculated as:
\begin{equation}
u_i^{(t)}=\frac{1}{N_p^{**}}\sum_{p=1}^{N_p^{**}}v^{(t)}_{i,p},\quad
T^{(t)}=\frac{1}{3N_p^{**}}\left(\sum_{p=1}^{N_p^{**}}\left|\bm{v}_p^{(t)}\right|^2-N_p^{**}\left|\bm{u}^{(t)}\right|^2\right).
\label{eq:temporary_statistic}
\end{equation}
To preserve the momentum and energy conservation, a linear transformation of particle velocities is employed. The updated particle velocity in $i$-th direction $v_i^{**}$ can be written as $v_i^{**}=\xi v_i^{(t)}+\eta_i$, with parameters:
\begin{equation}
\xi = \sqrt{\frac{T^{**}}{T^{(t)}}} \quad \text{and}\quad \eta_i = u_i^{**}-u_i^{(t)}\sqrt{\frac{T^{**}}{T^{(t)}}}.
\label{eq:linear_shifting}
\end{equation} 

In general, the replication and discarding procedures adjust the number of simulated particles to match the updated density predictions. The velocity scaling method then guarantees the mean velocities and temperature to align with the updated macroscopic properties. As the macroscopic synthetic equations inherently ensure mass, momentum, and energy conservation, when the macroscopic properties in the entire system reach their updated values, the conservation laws are still satisfied.

\subsubsection{Coupled sampling process in AAP-TRMC}

As shown in the last term of Eq.~\eqref{eq:AAPTRMC_collsion_operator}, in AAP-TRMC method, partial simulated particles are resampled from the distribution $f_M+\theta f_\alpha$, which is constructed from local macro variables. In the proposed GSIS method, this implement is coupled with the solution of the macroscopic synthetic equations. Therefore, Eq.~\eqref{eq:AAPTRMC_collsion_operator} is rewritten as,
\begin{equation}\label{eq:couplingtrmc}
f^{n+1}=\sum_{k=0}^mA_kf_k+A_{m+1}\left(f_M+\theta f_\alpha\right)|_{\rho^{**},\bm{u}^{**},T^{**}}.
\end{equation}
Note that $f_M+\theta f_\alpha$ is calculated based on the updated macroscopic properties. Since the moment and energy are preserved during the collision, the particle velocities scaling method is applied again in each iteration.

\subsection{General algorithm}
\begin{figure}[!t]
	\centering
	\includegraphics[width=0.7\textwidth]{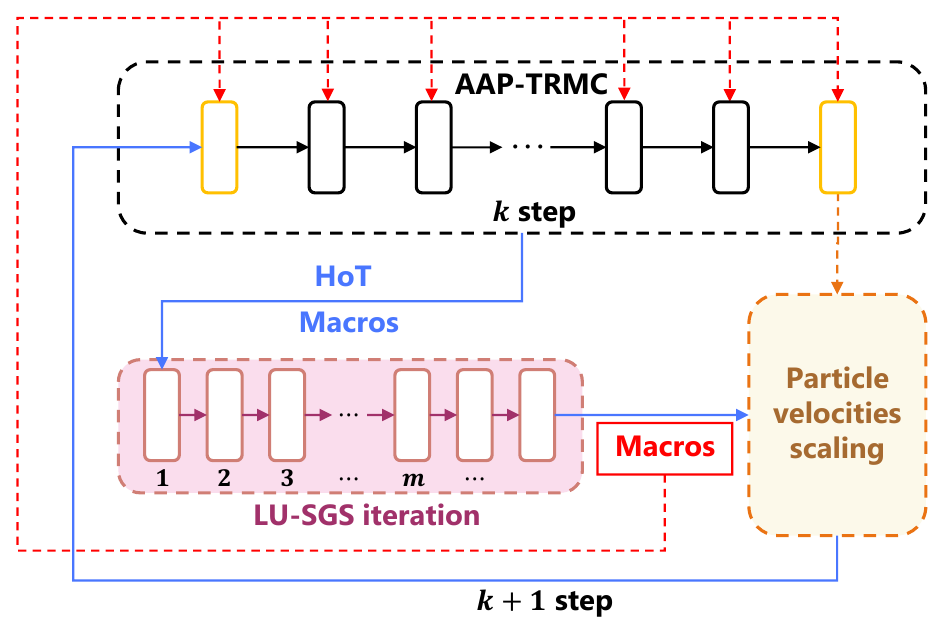}
	\caption{Flowchart of the particle-based GSIS algorithm. The black dashed box represents the original AAP-TRMC framework, and the macroscopic properties obtained from the synthetic equations will affect sampling process during the next sampling interval.}
	\label{flowchart}
\end{figure}
The general flowchart of the proposed coupling method is given in Fig.~\ref{flowchart}. In order to obtain accurate results, at least a second order interpolation in space should be used in this coupling method. That means the velocities of resampled particles are obtained by macroscopic properties as well as their derivatives at the particle location $\bm{x}$. Also, to improve the accuracy in time, the Strang splitting scheme~\cite{strang-1968} is applied. The overview of the coupling method is briefly outlined in Algorithm~\ref{algo:GSIS_procedure}.

\begin{algorithm}[!t]
    \caption{Overall algorithm of GSIS-AAPTRMC coupling method} 
    \label{algo:GSIS_procedure}
    \begin{algorithmic}[1]
        \Require
            Initial macroscopic properties $\bm{M}$;
        \Ensure
            Time-averaged macroscopic properties $\bm{M}$ after steady state;
        \State Draw certain number of particles within each cell according to initial $\bm{M}$ and Maxwellian distribution;
        \State Set $steps = 1$;
        \While {$steps \le \text{MaxSteps}$}
            \State First half advection: $\bm{x}\leftarrow\bm{x}+\frac{1}{2}\bm{v}\Delta t$ and diffuse boundary condition;
            \State Full collision: \Call{Collision}{particles, $\Delta t$};
            \State Another half advection: $\bm{x}\leftarrow\bm{x}+\frac{1}{2}\bm{v}\Delta t$ and diffuse boundary condition;
            \State Sampling: macroscopic properties sampled according to Eq.~\eqref{eq:statisticmacro};
            \If {$\text{MOD}\,(steps, N_s) == 0$};
            \State Extract time-averaged $\bm{M}^*$ (Eq.~\eqref{eq:statisticmacro}) and $HoT^*$ (Eq.\eqref{eq:highorderterms}) within interval $N_s$;
            \State Solve macroscopic synthetic equations and obtain $\bm{M}^{**}$, see section~\ref{sec:4};
            \State Obtain temporary macroscopic properties $\bm{M}^{(t)}$;
            \State Replicating and discarding particles
            \State Particle velocities scaling;
            \EndIf
            \State $steps ++$;
        \EndWhile
        \State
    \Function{Collision}{particles, $\Delta t$}
        \State Compute the upper bound $\bar{\sigma}=\left(\sigma_Tc_r\right)_{\max}$;
        \State Set $\tau = 1-\text{exp}(-\rho\bar{\sigma}\Delta t/\epsilon)$;
        \State Obtain different probabilities $A_0$, $A_1$, $A_2$, $A_3$ according to Eq.\eqref{eq:weightfunction}; 
        \State Set $N_1=\text{Iround}(A_1N/2)$ and perform normal collisions $f_1 = P(f^n,f^n)/\beta$;
        \State Set $N_2=\text{Iround}(A_2N/4)$ and perform collisions $f_2=P(f^n,f_1)/\beta$;
        \State Resample $N_R=\text{Iround}(A_3N)$ particles;
        \State Resample particle velocities based on Eq.~\eqref{eq:couplingtrmc};
        \State Recover moment and energy to their pre-collision states using particle velocities scaling method;
    \EndFunction
    \end{algorithmic}
\end{algorithm}
 \section{Numerical schemes of macroscopic solver}\label{sec:4}

\subsection{Finite-volume scheme for the macroscopic synthetic equations}

The macroscopic synthetic equations can be considered as conventional NS equations with HoTs as the source terms, therefore they can be solved efficiently by any CFD techniques. With unconstructed grids, the time-implicit finite volume method is employed, while standard Lower-Upper Symmetric Gauss-Seidel (LU-SGS) is used for the iterative process.

Integrating Eq.~\eqref{eq:Navior-Stokes} in a control volume $V$ and applying the Gauss theorem, the governing equation for macroscopic properties can be written as:
\begin{equation}
\frac{\partial}{\partial t}\int_V \bm{W}dV+\oint_S\left(\bm{F}_c+\bm{F}_v\right)d\bm{S}=-\oint_{S}\bm{F}_v^{HoT}d\bm{S},
\label{eq:integrating_NS}
\end{equation}
where $\bm{W}$ denotes the macroscopic variables and $\bm{F}_c, \bm{F}_v$ are the vectors of convective and viscous fluxes, which are given by (here we focus on the two-dimensional flows):
\begin{equation}
\bm{W} = \begin{bmatrix}\rho \\ \rho u_x \\ \rho u_y \\ \rho E \end{bmatrix},\ \ \bm{n}\cdot\bm{F}_c = \begin{bmatrix}\rho u_n \\ \rho u_xu_n+pn_x \\ \rho u_yu_n+pn_y \\ \rho Eu_n+pu_n \end{bmatrix},\ \ \bm{n}\cdot\bm{F}_v = \begin{bmatrix} 0 \\ \sigma_{xx}n_x+\sigma_{xy}n_y \\ \sigma_{xy}n_x+\sigma_{yy}n_y \\ \Theta_xn_x+\Theta_yn_y\end{bmatrix}.
\label{eq:expression_fluxes}
\end{equation}
Note that in  Eq.~\eqref{eq:expression_fluxes}, $\Theta_x = u_x\sigma_{xx}+u_y\sigma_{xy}+q_x$, $\Theta_y = u_x\sigma_{xy}+U_y\sigma_{yy}+q_y$, $u_n = \bm{u}\cdot\bm{n}$ and $\bm{n}$ is the unit normal vector of $d\bm{S}$. In the RHS of Eq.~\eqref{eq:expression_fluxes}, the viscous flux term depends on the HoTs in shear stress and heat flux $\bm{F}_v^{HoT}=\bm{F}_v(\bm{\sigma}^*,\bm{q}^*)-\bm{F}_v(\bm{\sigma}_{NS}^*,\bm{q}_{NS}^*)$. Under the framework of the finite volume method, the time-implicit discretized form of the governing equation \eqref{eq:expression_fluxes} can be written as:
\begin{equation}
\frac{\bm{W}_i^{n+1}-\bm{W}_i^{n}}{\Delta t}+\frac{1}{V_i}\sum_{j\in N(i)}\bm{n}\cdot\left(\bm{F}_{ij}^{n+1}+\bm{F}_{v,ij}^{HoT}\right) S_{ij}=0,
\label{eq:time_implicit}
\end{equation}
where the flux $\bm{F}_{ij}=\bm{F}_{c,ij}+\bm{F}_{v,ij}$ includes both convection and viscous terms. Note that the high-order terms on the right-hand-side in Eq.~\eqref{eq:time_implicit} remain constant throughout since they are directly extracted from the particle-based scheme. By introducing the incremental variables $\Delta \bm{W}_i^m=\bm{W}_i^{m+1}-\bm{W}_i^m$ and $\bm{F}_{ij}^{m+1} = \bm{F}_{ij}^m+\Delta\bm{F}_{ij}^m$ with $m$ standing for the inner iteration index when solving macroscopic equations,we obtain:
\begin{equation}
\frac{1}{\Delta t_i}\Delta \bm{W}_i^m+\frac{1}{V_i}\sum_{j\in N(i)}\bm{n}\cdot\Delta \bm{F}_{ij}^{m}S_{ij}=-R_i^m-R_i^{HoT},
\label{eq:time_implicit_1}
\end{equation}
where $\Delta t_i$ is the pseudo time step and $R$ stands for the residues of fluxes based on the NS relations and high order terms:
\begin{equation}
\begin{aligned}
R_i^m &= \sum_{j\in N(i)}\bm{n}\cdot\left[\bm{F}_{c}^{m}+\bm{F}_v\left(\bm{\sigma}_{NS}^m,\bm{q}_{NS}^m\right)\right]_{ij}S_{ij},\\
R_i^{HoT}&=\sum_{j\in N(i)}\bm{n}\cdot\left(\bm{F}_v^{HoT}\right)_{ij}S_{ij}.
\end{aligned}
\label{eq:fluxes_NS_HOT}
\end{equation}
Note that the general form of the fluxes at the cell face can be represented as $F_{ij}=F(\bm{W}_L,\bm{W}_R,S_{ij})$, where $\bm{W}_{L/R}$ stands for the reconstruction of the macroscopic properties in two different cells adjacent to this cell face: $\bm{W}_{L/R}=\bm{W}_{i/j}+\phi\nabla(\bm{W}_{i/j}\cdot \bm{x}_{i/j})$. Several reconstruction methods such as Rusanov scheme \cite{sod-1978} and Roe scheme \cite{roe-1981} can be applied to enhance the numerical stability. Here, we recommend using the AUSMPW \cite{kim-2001} scheme for cases involving extremely high Mach numbers. 

In general, the calculation of macroscopic implicit fluxes in left-hand-side of Eq.~\eqref{eq:time_implicit_1} is approximated by the first-order flux in the Euler equation, which gives:
\begin{equation}
\Delta\bm{F}_{ij}^m = \frac{1}{2}\left[\Delta\bm{F}_i^m+\Delta\bm{F}_j^m+\Gamma_{ij}\left(\Delta W_i^m-\Delta W_j^m\right)\right],
\label{eq:Eulerfluxes}
\end{equation}
where the parameter $\Gamma_{ij}$ represents the spectral radius in Jacobian matrix of Euler fluxes. Consider the viscous effect, the stabilization term $\Gamma_{\nu}$ related to the dynamic viscosity should be added, which gives:
\begin{equation}
\Gamma_{ij}=\left(u_n+a_s\right)+\Gamma_{\nu}= \left(u_n+a_s\right)+\frac{2\mu}{\rho\left|\bm{n}_{ij}\cdot(\bm{x}_j-\bm{x}_i)\right|},
\end{equation}
where $a_s$ is the speed of sound. Since the conservation law is satisfied for the geometrically enclosed finte volume cell, the interface fluxes through $i$-th cell should collectively sum to zero, given by $\sum_{j\in N(i)}\bm{F}_iS_{ij}=0$. Thus, the general implicit governing equations for macroscopic properties in Eq.~\eqref{eq:time_implicit_1} can be expressed as:
\begin{equation}
\left(\frac{1}{\Delta t_i}+\frac{1}{2V_i}\sum_{j\in N(i)}\Gamma_{ij}S_{ij}\right)\Delta \bm{W}_i^m+\frac{1}{2V_i}\sum_{j\in N(i)}\left(\Delta \bm{F}_j^m-\Gamma_{ij}\Delta \bm{W}_j^m\right)S_{ij}=-R_i^m-R_i^{HoT},
\end{equation}
which can be solved by the classical Lower Upper Symmetric Gauss-Seidel (LU-SGS) iteration technique, which is widely applied in computational fluid dynamics. For the sake of simplicity, the details of LU-SGS iteration technique are not further elaborated here, and specific settings for the LU-SGS process can be found in Ref.~\cite{zhu-2021}.

\subsection{Boundary treatment and convergence criteria for macroscopic solver}

In addition to extracting macroscopic properties and HoTs during a specific time-averaged sampling interval $N_s$, the synthetic equation solver requires information on boundary fluxes as the initial input. In the previous study \cite{zhu-2021}, the macroscopic equations were just solved in the inner domain and fluxes through the first four cell layers adjacent to the wall boundary were held constant. That means during the iteration process, a fixed boundary flux is provided while the fluxes are updated within the inner domain, which will become unstable for problems with incompatible boundary conditions. This problem is solved in Ref.~\cite{zeng-2023}, which provides a method similar to Roe scheme to modify the interface fluxes and update the boundary fluxes in each iteration step. Later, a generalized boundary treatment has been developed, which converges even faster~\cite{liu-2024}. In this method, the linear constitutive relations with high order terms are applied to construct the VDF near the surface based on the framework of Grad 13 moment method. The VDF for monatomic gas on the left and right $(L,R)$ sides of the interface $i,j$ is reconstructed as follows,
\begin{equation}
f_{ij}^{L,R}=f_M\left[1+\frac{c_{<i}c_{j>}\left(\sigma_{ij,NS}+\text{HoT}_{\sigma_{ij}}\right)}{2pT}+\frac{2c_k\left(q_{k,NS}+\text{HoT}_{q_k}\right)}{5pT}\left(\frac{c^2}{2T}-\frac{5}{2}\right)\right].
\label{eq:G13_interface}
\end{equation}
Thus, the macroscopic flux at the interface $i,j$ can be obtained:
\begin{equation}
    \bm{F}_{ij}=
    \begin{bmatrix}
    \rho u_n \\ \rho u_xu_n \\ \rho u_yu_n \\ \rho Eu_n 
    \end{bmatrix}
    =
    \begin{bmatrix}
    \int_{v_n>0}v_nf_{ij}^{L}d\bm{v}+\int_{v_n<0}v_nf_{ij}^{R}d\bm{v} \\ 
    \int_{v_n>0}v_xv_nf_{ij}^{L}d\bm{v}+\int_{v_n<0}v_xv_nf_{ij}^{R}d\bm{v} \\
    \int_{v_n>0}v_yv_nf_{ij}^{L}d\bm{v}+\int_{v_n<0}v_yv_nf_{ij}^{R}d\bm{v} \\
    \int_{v_n>0}\frac{1}{2}\bm{v}^2v_nf_{ij}^{L}d\bm{v}+\int_{v_n<0}\frac{1}{2}\bm{v}^2v_nf_{ij}^{R}d\bm{v}
    \end{bmatrix}
    .
    \label{eq:flux_G13}
\end{equation}

Additionally, the convergence criteria for the macro solver will affect the efficiency and stability of the whole algorithm. The error of the conservative variables between two successive steps can be defined as
\begin{equation}
e_k = \sqrt{\frac{\sum_i\left(\psi_i^k-\psi_i^{k-1}\right)^2V_i}{\sum_i\left(\psi_i^{k-1}\right)^2V_i}} < \varepsilon_{in},
\label{eq:errordefinition}
\end{equation}
where $\psi_i^k$ represents the conservative properties in $\bm{W}$ in $i$-th cell and $k$-th iteration step. The convergence criteria $\varepsilon_{in}$ is set to an exceedingly small value in most deterministic methods. However, due to the inevitable fluctuations of conservative properties, it is unnecessary to enforce a extremely small value for $\varepsilon_{in}$. Empirically, setting the value of $\varepsilon_{in}$ to $10^{-4}$ generally suffices in most cases, depending on the scale of fluctuations for conservative properties.



 \section{Numerical results}\label{sec:5}

The argon gas is considered for all simulations in this section, incorporating the VHS model in collision and Maxwellian diffuse boundary condition at the solid surface. The Knudsen number is defined as:
\begin{equation}
    \text{Kn} = \frac{\mu_{0}}{p_{0} L_{ref}}\sqrt{\frac{\pi k_B T_{0}}{2m}},
\end{equation}
where $L_{ref}$ represents the characteristic length for different systems. $p_0$ represents the reference pressure under the condition of reference number density $n_0$ and temperature $T_0$. The viscosity is calculated as $\mu=\mu_{0}(T/T_{0})^\omega$ with the exponent $\omega = 0.81$, while the reference viscosity $\mu_{0}$ is obtained at the reference temperature $T_0$.


As discussed in previous section, GSIS requires a moderate sample size $N_s$ to obtain the macroscopic properties. If $N_s$ is large, the synthetic equations are solved after lengthy periods of particle evolution, which consequently diminishes the acceleration effect. Conversely, if $N_s$ is small, the fluctuations in macroscopic properties will lead to the instability of the algorithm. Thus, an empirical value of $N_s=100$ is chosen for all simulations.



\subsection{Planar Fourier flow}

Consider the argon gas between two parallel plates located at $x_L=0$ and $x_R=1$ with the temperature $T_L=0.5$ and $T_R=1.5$, respectively. 
When the steady state is reached, $\partial/\partial t=0$, $\partial/\partial y = 0$, $\bm{u}=0$, $q_y=0$, and $\sigma_{xy}=\sigma_{yx}=0$. 
Thus, the macroscopic synthetic equations are simplified to: 
\begin{equation}\label{s3}
	\begin{aligned}
		&q_x = -\kappa\frac{\partial T}{\partial x}+\text{HoT}_{q_{x}}=-\kappa\frac{\partial T}{\partial x}+\frac{1}{2}\int f^*v^{*2}v_{x}^*d\vec{v}+\kappa^*\frac{\partial T^*}{\partial x},\\
   &\frac{\partial q_x}{\partial x}=0,
	\end{aligned}
\end{equation}
and 
\begin{equation}\label{s1}
	\begin{aligned}
		&\sigma_{xx} = \text{HoT}_{\sigma_{xx}}=\int f^*\left(\frac{2}{3}v_x^{*2}-\frac{1}{3}v_y^{*2}-\frac{1}{3}v_z^{*2}\right)d\vec{v},\\
  &\frac{\partial}{\partial x}\left(p+\sigma_{xx}\right)=0, \\
	&	p=\rho T, 
	\end{aligned}
\end{equation}
where the variables with superscript * are the time-averaged results in the previous 100 steps (note that 1000 particles are initially allocated in each cell to reduce the thermal fluctuations).
According to Eq.~\eqref{s3}, the heat flux across the whole system is a constant. 
Thus, by choosing the temperature in the first and last cell, the heat flux and temperature in the bulk region can be obtained by solving the second equation. Similarly, with the normal stress $\sigma_{xx}$ obtained from the AAP-TRMC, the pressure and density can be obtained by solving the 
second and third equations in Eq.~\eqref{s1}, respectively. 

\begin{figure}[!t]
	\centering
        \subfloat[]{\includegraphics[width=0.45\textwidth]{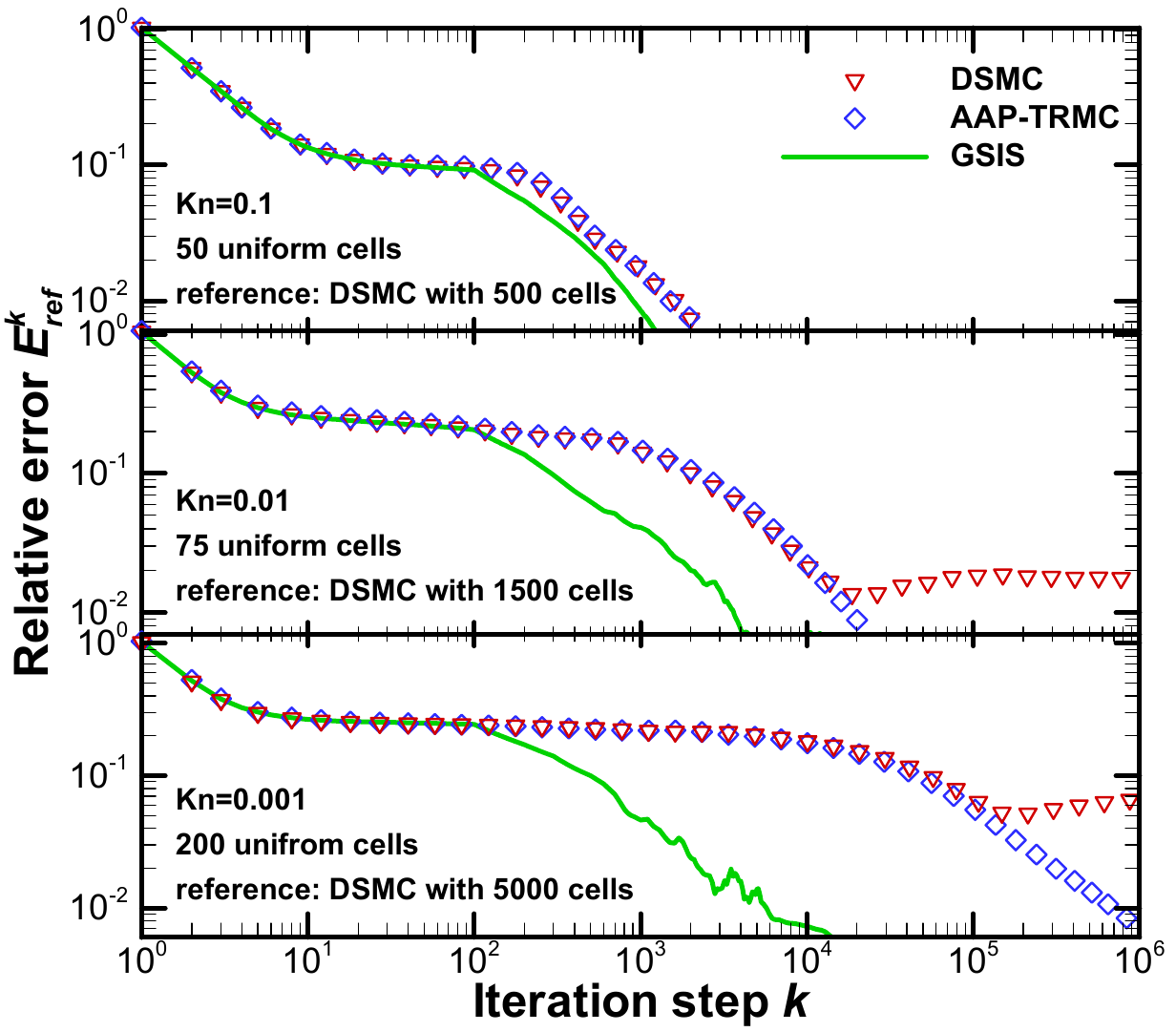}\label{Convergence_VALI_a}}
        \subfloat[]{\includegraphics[width=0.48\textwidth]{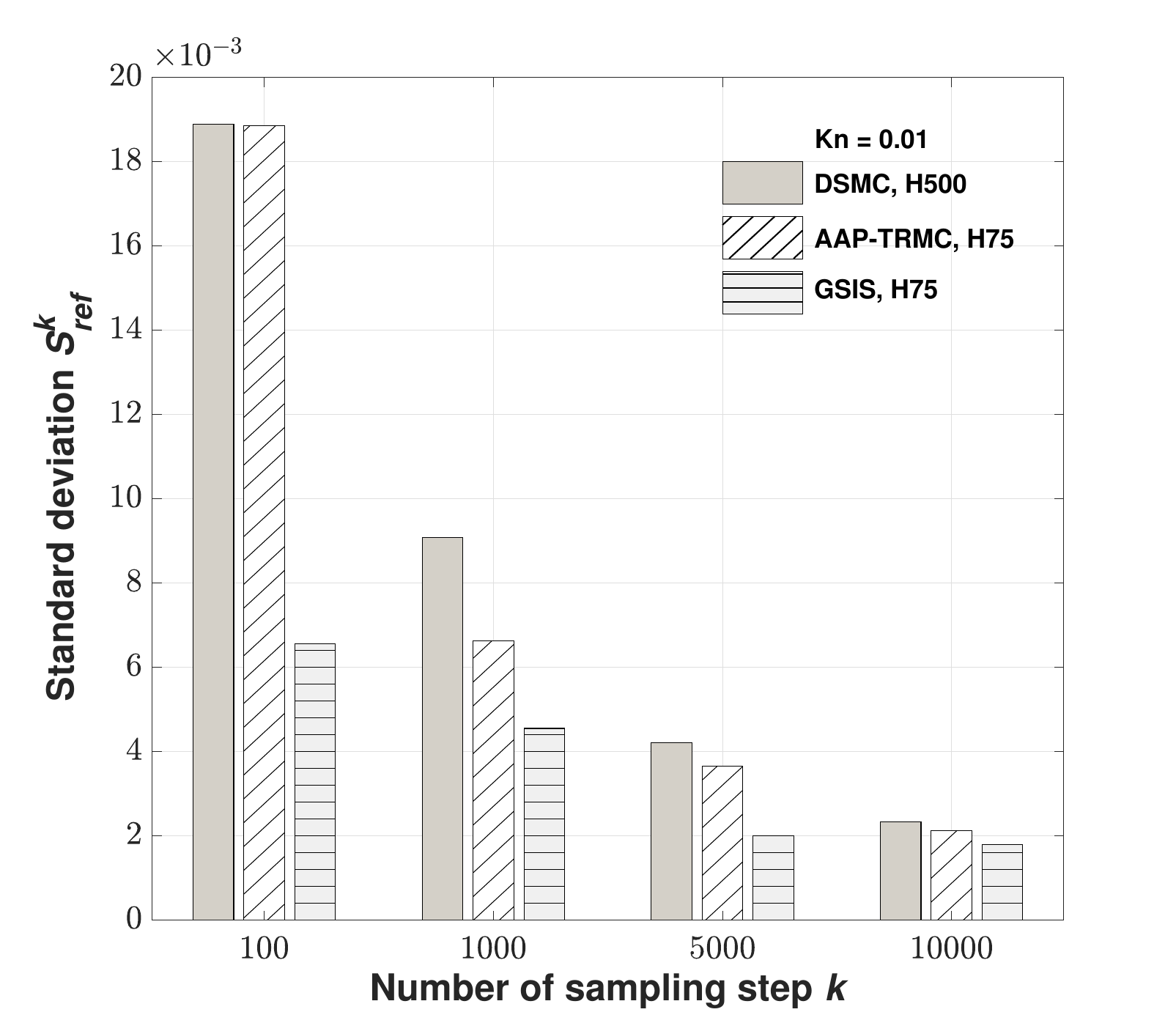}\label{Convergence_VALI_b}}
	\caption{(a) Evolution of the relative error~\eqref{eq:relative_error} in the planar Fourier flow, when $\text{Kn}=0.1$, 0.01, and 0.001. (b) Evolution of the standard deviation~\eqref{eq:relative_fluctuations} during the steady state for $\text{Kn}=0.01$, where the sampling starts when the steady state is reached. H75 means that the computational domain is discretized uniformly by 75 cells. In all numerical simulations, the CFL number is 0.2. }
	\label{Convergence_VALI}
\end{figure}

To assess how quickly the numerical schemes evolve towards the steady state, we define the relative error as follows:
\begin{equation}\label{eq:relative_error}
E^k_{ref}=\max\left(\sum_i{\left|\frac{\bar{\rho}_i^k-\rho^{ref}_{i}}{\rho^{ref}_{i}}\right|\Delta x},\,\sum_i{\left|\frac{\bar{T}_i^k-T_i^{ref}}{T_i^{ref}}\right|\Delta x}\right),
\end{equation} 
where the reference solutions $\left(\rho^{ref},\,T^{ref}\right)$ are obtained from the original DSMC with spatial cell size less than one fifth of the mean free path.
Note that  $\bar{\rho}^k$ and $\bar{T}^k$ are not the macroscopic quantities at the $k$-th step, but are time-averaged from the beginning of the computation to the $k$-th step. Next, to evaluate the criteria for terminating the computational simulation during the steady state, we define the standard deviations describing the temporary fluctuations as follows:
\begin{equation}\label{eq:relative_fluctuations}
S^k_{ref}=\max\left(\sqrt{\sum_i{\left(\Tilde{\rho}^k_i-\Tilde{\rho}^{ref}_{i}\right)^2\Delta x}},\,\sqrt{\sum_i{\left(\Tilde{T}^k_i-\Tilde{T}^{ref}_{i}\right)^2\Delta x}}\right),
\end{equation} 
where the tilde "$\sim$" represents the time-averaged values sampled $k$ steps after the flow field reaching the steady state. For the reference solutions, i.e., $\Tilde{\rho}^{ref}$ and $\Tilde{T}^{ref}$, $k=10^6$ sampling steps are calculated.


As depicted in Fig.~\ref{Convergence_VALI_a}, when Kn = 0.1, the DSMC, AAP-TRMC and GSIS utilizing identical spatial discretization show similar convergence rates. However, as the Knudsen number diminishes, GSIS demonstrates its higher computational efficiency over the other two. Specifically, at $\text{Kn}=0.001$, when only 200 uniform cells are employed, DSMC experiences significant numerical dissipation. Conversely, AAP-TRMC and GSIS, leveraging their robust AP property, deliver accurate outcomes at the same number of spatial cells, e.g., when the spatial cell size is about 5 times larger than the molecular mean free path. 
Strikingly, GSIS introduces a pronounced inflection in the convergence curve during its initial solving of the macroscopic synthetic equations and subsequent guidance of the flow field. The sustained guidance over the next 100 iterations markedly expedites the flow field's evolution. Ultimately, GSIS reduces the iteration number by two-order-of-magnitude when compared to the AAP-TRMC in the transition state. Figure~\ref{Convergence_VALI_b} shows the standard deviation for different number of sampling steps, which represents the fluctuations of the time-averaged macroscopic properties during the steady state. When $\Kn=0.01$, the standard deviation for GSIS decreases to 0.2\% after 5000 sampling steps.
Since the solutions to the synthetic equations continuously guide the subsequent evolution in GSIS, time-averaged samples in the steady state can be reduced.
Thus, to attain a comparable level of fluctuation, the other two algorithms requires a substantially higher number of sampling steps, specifically around 10,000.

\begin{table}[!t]
 \centering
 \caption{\label{tab1}Computational overhead of the DSMC, AAP-TRMC, GSIS (the upper, middle, and lower rows of each Knudsen number, respectively) for the planar Fourier flow at different Knudsen numbers. Simulations denoted with a superscript * were performed by the Stochastic PArallel Rarefied-gas Time-accurate Analyzer (SPARTA, \url{https://sparta.sandia.gov/}) using 80 cores, while other simulations are done by 4 cores of Intel(R) Core(TM) i7-10700K CPU @ 3.80GHz processor. The computational time is given in core$\cdot$hours. }
\begin{threeparttable} 
  \begin{tabular}{ c c  c c c c}\toprule
 \multirow{2}{*}{Kn}  &  \multirow{2}{*}{$\Delta x$} &  \multicolumn{2}{c}{Transition state}  &  \multicolumn{2}{c}{Steady state} \\ \cmidrule(r){3-4} \cmidrule(r){5-6}
    ~                & ~                                 & steps&time&steps   &    time\\ \hline        
 \multirow{3}{*}{1.0} & 0.02$\lambda$&  1000   &    0.004   &          5000&      0.02     \\
                        & 0.02$\lambda$  &  1000      &    0.004   &     5000       &      0.03     \\
                        &  0.02$\lambda$  &  800   &    0.005   &     5000       &      0.03     \\ 
                        \addlinespace
\multirow{3}{*}{0.1}  & 0.2$\lambda$ & 600  &    0.003   &    5000     &      0.03     \\
 ~                    & 0.2$\lambda$                             & 600  &    0.004   &    5000     &      0.04     \\
 ~                    &0.2$\lambda$                                 & 300  &    0.002   &    5000     &      0.04     \\ \addlinespace
\multirow{3}{*}{0.01} & 0.2$\lambda$                 & 30000  &    1.27   &    10000     &      0.44\\
 ~                    & 1.67$\lambda$ & 6000  &    0.1   &    10000     &      0.18 \\
 ~                    & 1.67$\lambda$  & 800  &    0.02   &    5000     &      0.09 \\ \addlinespace
\multirow{3}{*}{0.001}& 0.33$\lambda$\tnote{*}              & $2\times10^6$ &    113   &    50000     &      8.1\\
 ~                    & 5$\lambda$ & $1.5\times10^5$  &    13   &    50000     &      4.2 \\
 ~                    & 5$\lambda$ & 1500  &    0.11   &    10000     &      0.79 \\ \addlinespace
\bottomrule
\end{tabular}
 \end{threeparttable}
\end{table}

The iteration steps and total CPU time for the three schemes are compared in Table~\ref{tab1}. 
Note that in this case, when the relative error $E_k^{ref}$ is below 4\%, the solution is deemed to reach the steady state. And after this, the sampling is terminated when the fluctuation of macroscopic properties, $S^k_{ref}$, reaches approximately 0.002.
When $\text{Kn}=0.1$ and $1.0$, GSIS shows little advantage in reducing the iteration step in the transition state and steady state, and due to the additional cost in solving the synthetic equations every 100 steps, the overall cost is larger than DSMC. However, as the Knudsen number decreases, despite the additional solving of macroscopic equations in GSIS, the computational time remains significantly lower than that for the resolution of kinetic equations. Consequently, the overall computational time is significantly reduced in GSIS. For example, when $\text{Kn}=0.001$, the CPU time is reduced by nearly two orders of magnitude compared to DSMC, and by approximately 20 times compared to AAP-TRMC. For lower Knudsen number cases, since GSIS can reduce the iteration step by several orders of magnitude, the additional cost in solving synthetic equations can be neglected. Thus, GSIS has its superior computational efficiency compared to other two schemes in the near-continuum regime.


\subsection{Lid-driven cavity flow}

The computational domain is a $L\times L$ square cavity and the Knudsen number is defined according to the width $L$. All solid walls have the same temperature $T_w=1$ (normalized with respect to $T_{0}$). The top lid of the cavity moves horizontally along x-axis at a speed of $U_w = \sqrt{2}$ (normalized by $c_0$) when $\Kn\ge 0.01$. And in the near-continuum regime, to avoid the turbulence, $U_w$ is varied to 0.21 and 0.42, corresponding to $\Kn=2.63\times10^{-3}$ and $5.26\times10^{-4}$, respectively. At the beginning, each computational cell is populated with 100 simulation particles, and the velocities of all particles are determined based on the Maxwell distribution function of density one, velcoity zero, and temperature one.

\begin{table}[!t]
 \centering
\caption{\label{tab2}
Computational overhead of the DSMC, AAP-TRMC, GSIS (the upper, middle, and lower rows of each Knudsen number, respectively) for the lid-driven flow. The computational time is given in core hours.
``-'' means that the computational cost of DSMC is unaffordable. 
}
\begin{threeparttable} 
  \begin{tabular}{c c  c c c c c c c}\toprule
 \multirow{2}{*}{Kn} &  \multirow{2}{*}{Re}  &  \multirow{2}{*}{$U_w$} & \multirow{2}{*}{CFL}   & \multirow{2}{*}{$N_{\text{cell}}$}  & \multicolumn{2}{c}{Transition state}  &  \multicolumn{2}{c}{Steady state} \\ \cmidrule(r){6-7} \cmidrule(r){8-9}
 ~ & ~ & ~ & ~ &~                                 & steps&time&steps   &    time\\ \hline
 \multirow{3}{*}{1.0} &  \multirow{3}{*}{1.77}  & \multirow{3}{*}{1.41} & \multirow{3}{*}{0.2}    &  \multirow{3}{*}{$50\times50$} &    300   & 0.009 &  $10^4$ &      0.28     \\
~   & ~   & ~     & ~            &    ~ &    300   & 0.016 &  $10^4$ &      0.57     \\
~   & ~   & ~     & ~            &    ~ &    300   & 0.018 &  $10^4$ &      0.74     \\  \addlinespace
 \multirow{3}{*}{0.1} & \multirow{3}{*}{17.73} & \multirow{3}{*}{1.41} &   \multirow{3}{*}{0.2}   &  \multirow{3}{*}{$50\times50$}  &    800   & 0.027 &  $10^4$ &      0.32     \\
~   & ~   & ~     & ~            &    ~ &    800   & 0.042 &  $10^4$ &      0.69     \\
~   & ~   & ~     & ~            &    ~ &    300   & 0.032 &  $10^4$ &      0.53     \\  \addlinespace
 \multirow{3}{*}{0.01}& \multirow{3}{*}{177.28} & \multirow{3}{*}{1.41} &  0.2   &  $500\times500$&  20000   & 17    &  $10^5$&      80        \\
~   & ~   &~ & 0.5 & $100\times100$   &   7000   & 3.2 &  $10^5$ &      41     \\
~   & ~   & ~     & 0.5            & $100\times100$ &    500   & 0.29 &50000&      27     \\  \addlinespace 
 \multirow{3}{*}{$2.63\times10^{-3}$}& \multirow{3}{*}{100}&  \multirow{3}{*}{0.21}  &-& -  &  -   & -    &  -&      -        \\
~   & ~      & ~ & 0.5   & $150\times150$ &   $>$30000   & $>$42 &  $>$50000 &      $>$72     \\
~   & ~   & ~     & 0.5            &  $150\times150$ &   1000   & 4.0 &  $10^4$&      31     \\  \addlinespace
 \multirow{3}{*}{$5.26\times10^{-4}$}& \multirow{3}{*}{1000}&  \multirow{3}{*}{0.42}  &-& -  &  -   & -    &  -&      -        \\
~   & ~      & ~ & 0.5   & $150\times150$ &   $>$40000   & $>$54 &  $>$50000 &     $>$76     \\
~   & ~   & ~     & 0.5            &  $150\times150$ &   2000   & 7.6 &  $10^4$&      35  
    \\  
\bottomrule
\end{tabular}
 \end{threeparttable}
\end{table}

\begin{figure}[t]
	\centering
\subfloat{\includegraphics[width=0.41\textwidth]{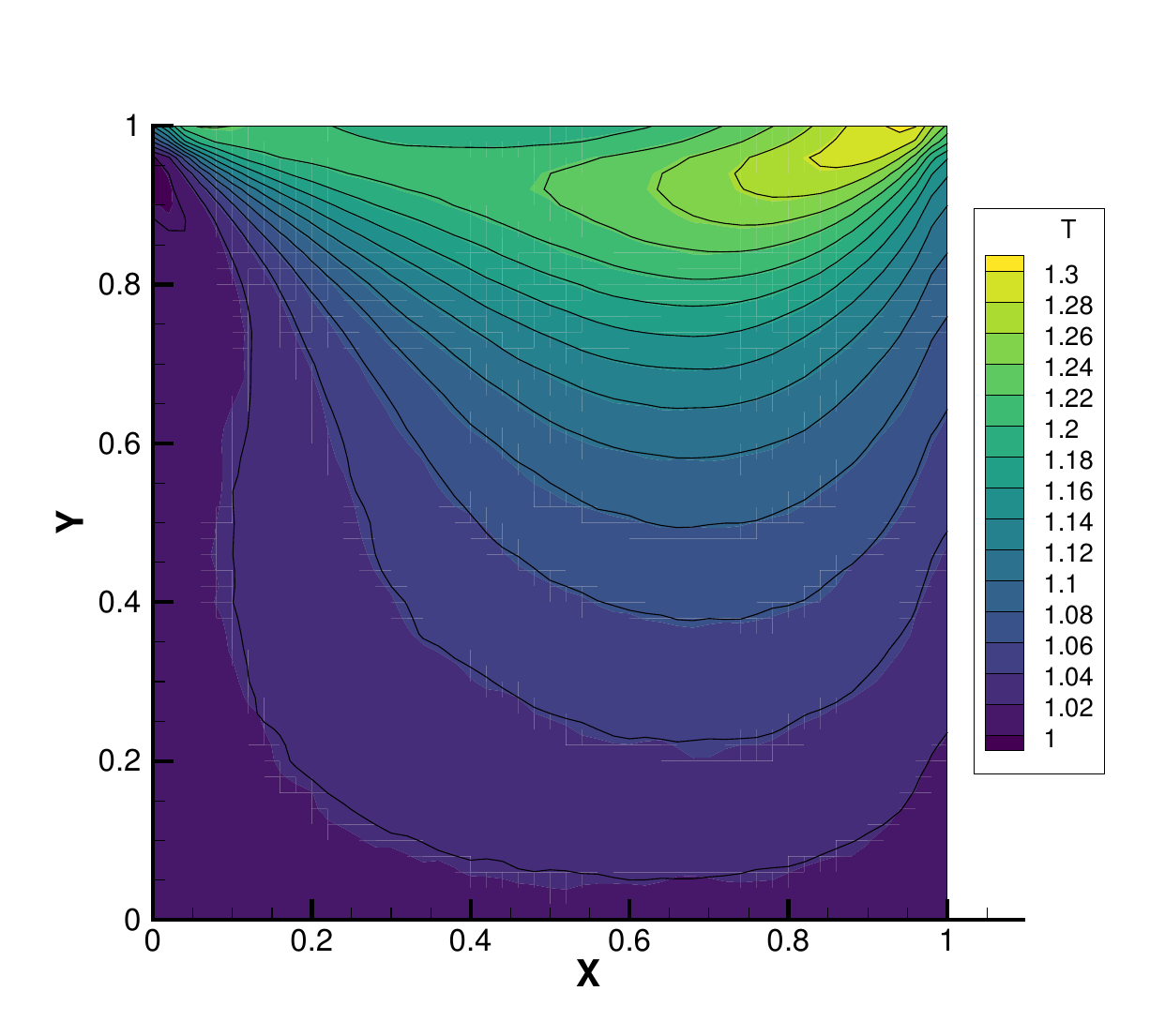}\label{Cavity-Tcontour-Kn01}}
	\subfloat{\includegraphics[width=0.41\textwidth]{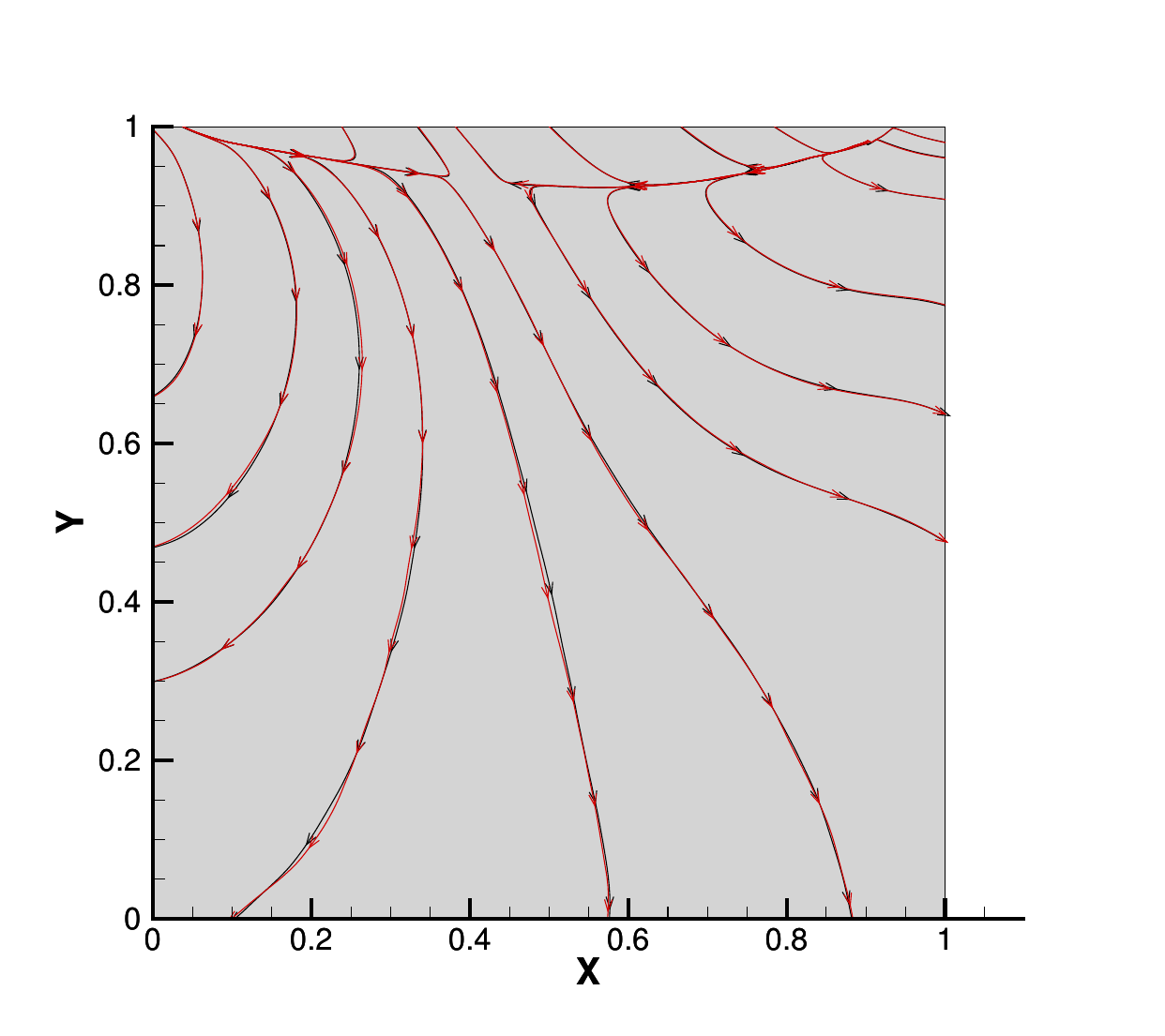}\label{Cavity-heatflux-Kn01}}\\
	\subfloat{\includegraphics[width=0.41\textwidth]{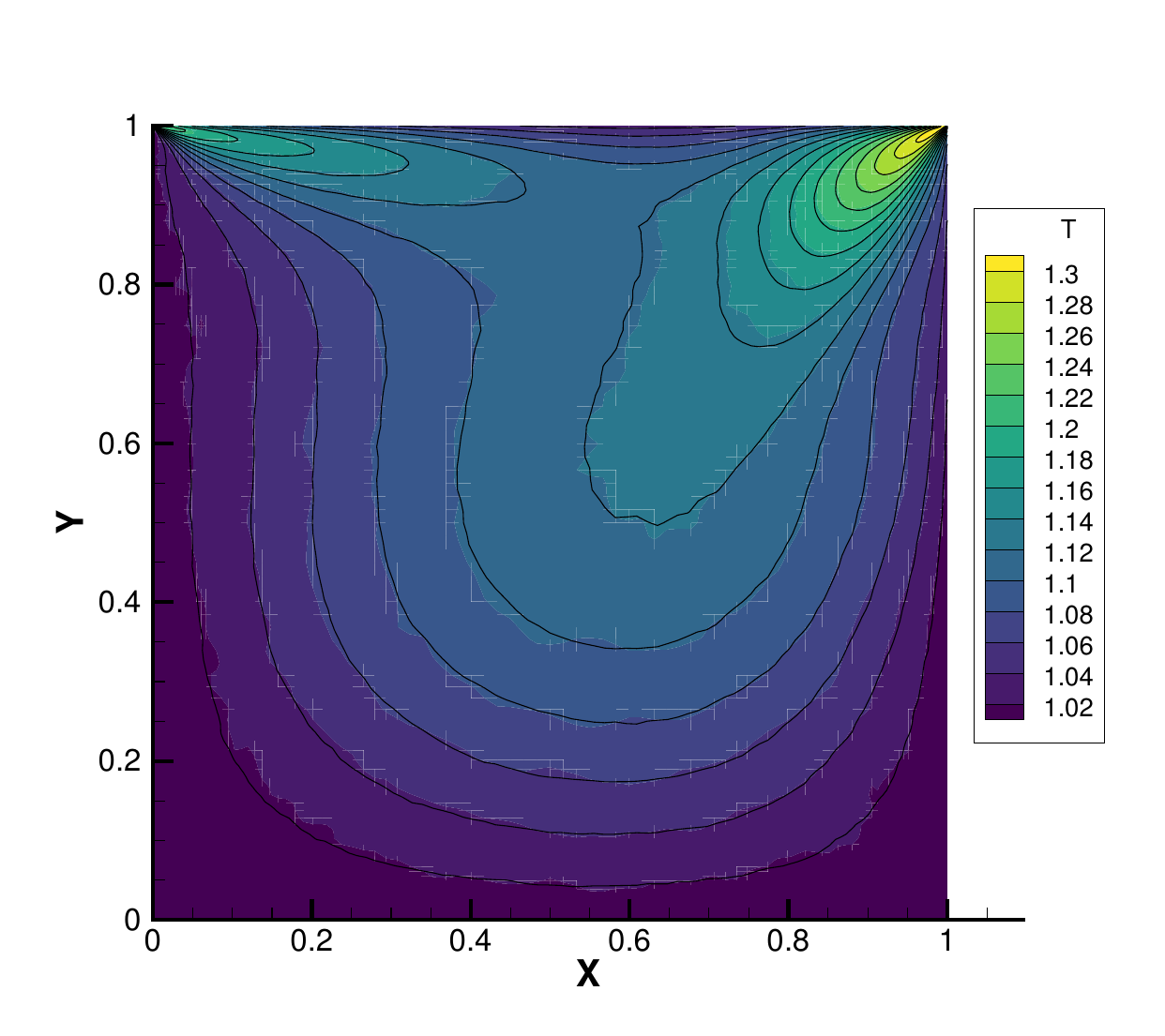}\label{Cavity-Tcontour-Kn001}}
	\subfloat{\includegraphics[width=0.41\textwidth]{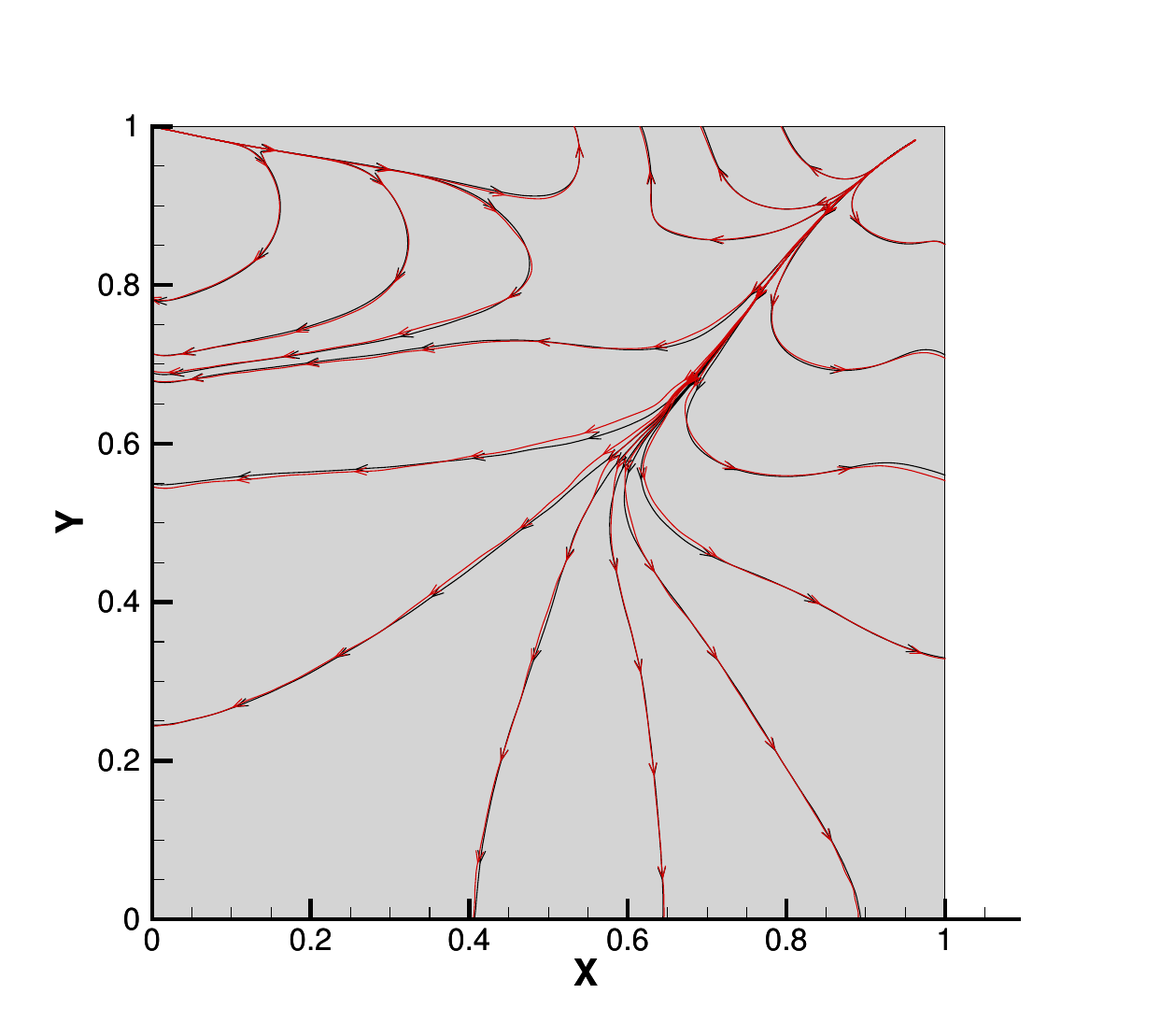}\label{Cavity-heatflux-Kn001}}
	\caption{Temperature contours (left) and heat flux streamlines (right) in the lid-driven cavity flow, when Kn=0.1 (top) and 0.01 (bottom). For temperature contours, the colored backgrounds (black lines) represent the results obtained by GSIS (DSMC), while for heat flux streamlines, GSIS (DSMC) results are denoted by red (black) lines.}
	\label{Cavity_Tq_figure}
\end{figure}




Figure~\ref{Cavity_Tq_figure} compares the temperature and the heat flux streamlines for different schemes, which demonstrates the accuracy of GSIS in the slip flow regime. When $\text{Kn}=0.1$, $50\times50$ uniform grids are employed for both schemes. When $\text{Kn}=0.01$, $100\times100$ non-uniform grids are applied, which are refined near the solid walls, e.g., the first layer cell has a thickness of $\Delta x_{\text{min}}=0.005$. In comparision, $500\times500$ uniform grids are applied in DSMC. The results show that the GSIS and DSMC results match well with each other.



\begin{figure}[!t]
	\centering
	\includegraphics[width=0.45\textwidth]{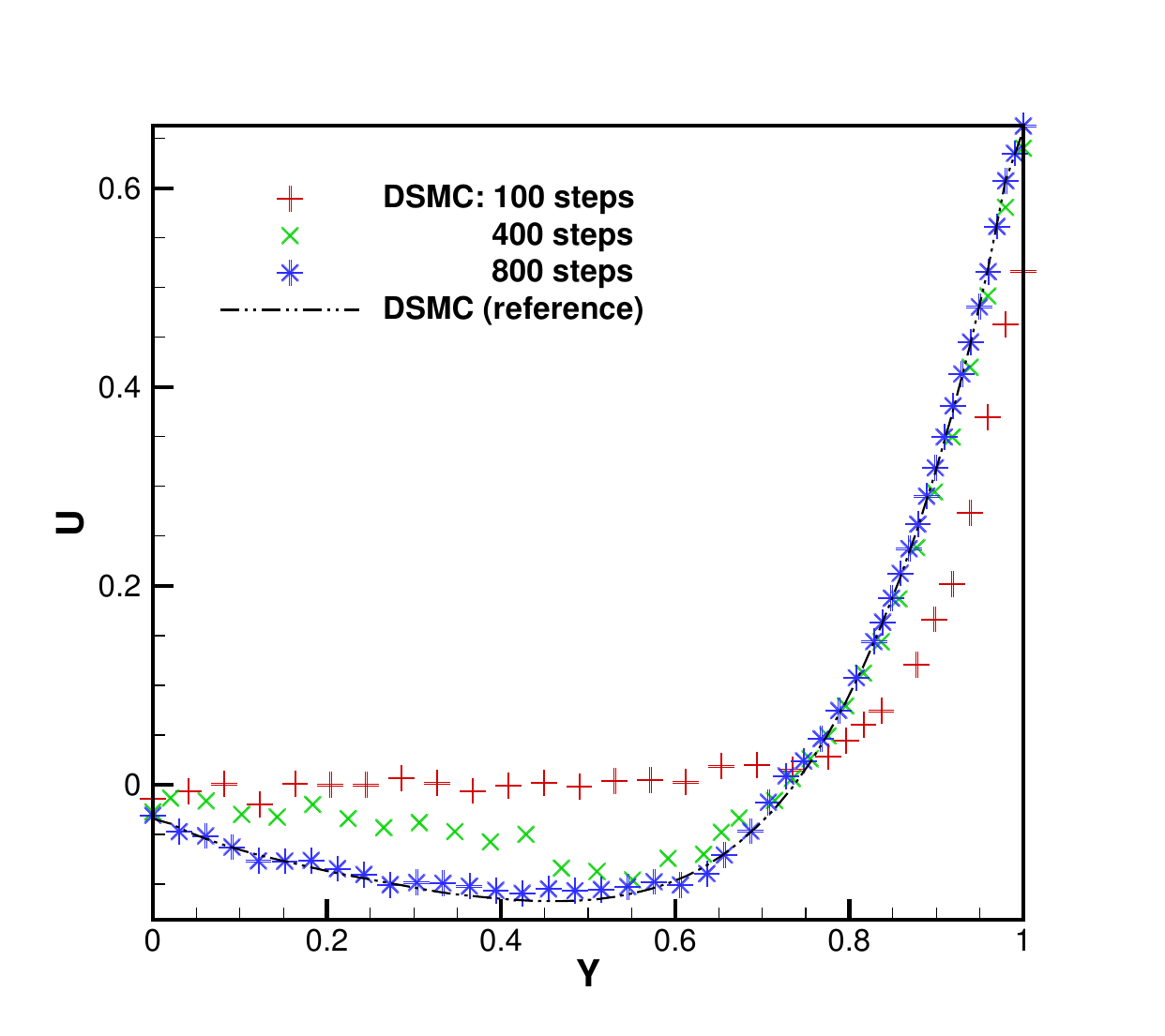}
	\includegraphics[width=0.45\textwidth]{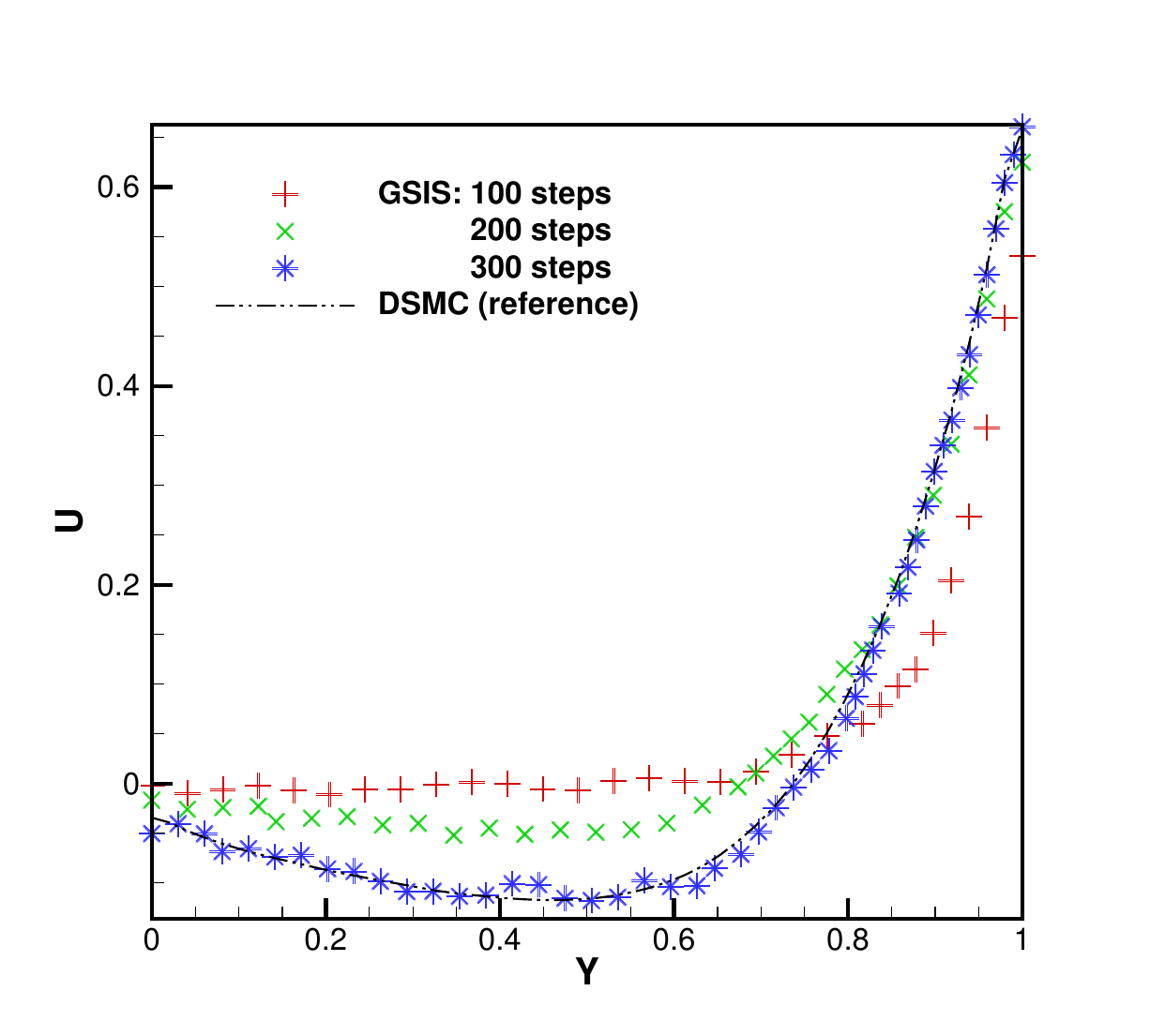}\\
	\includegraphics[width=0.45\textwidth]{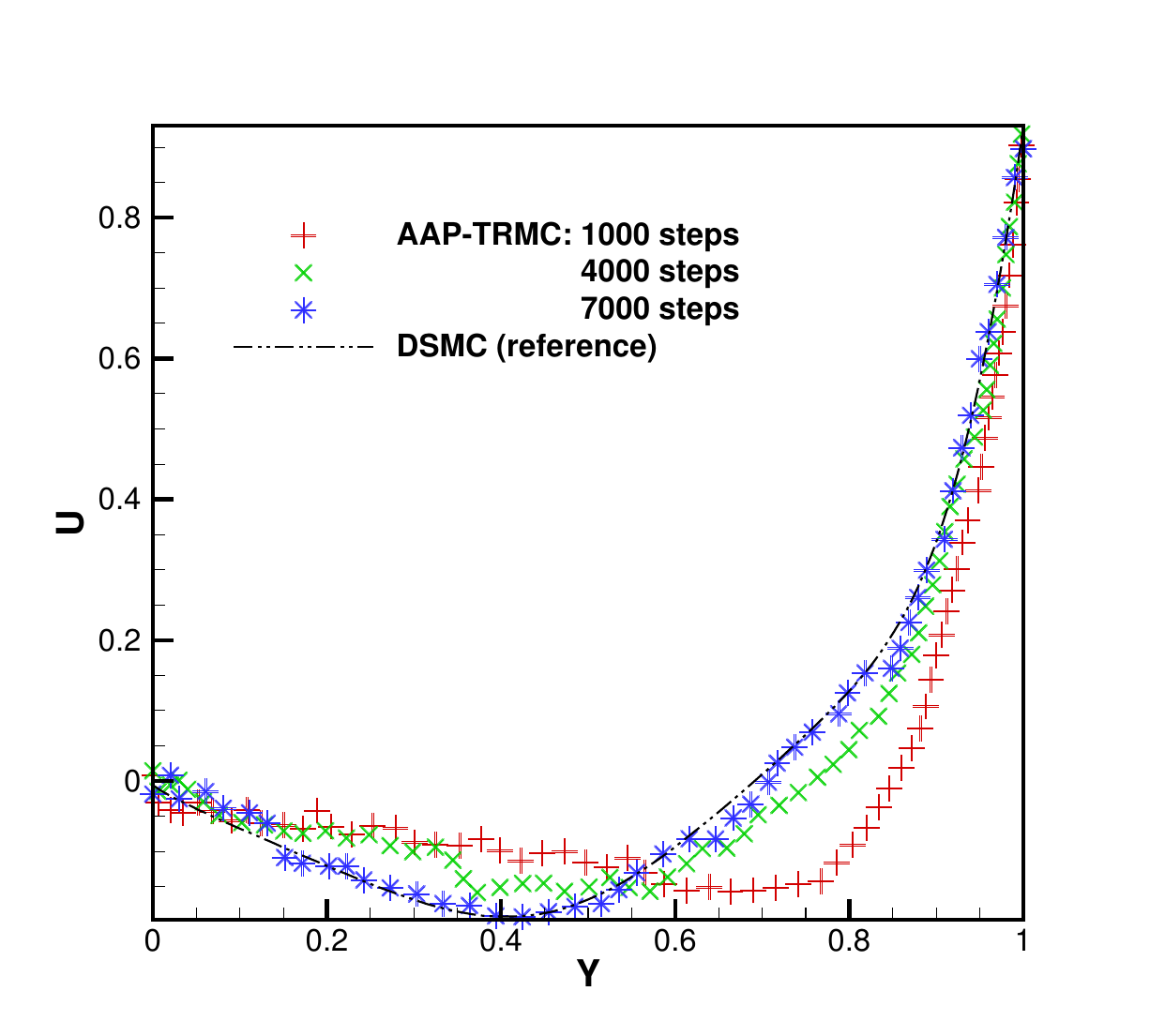}
	\includegraphics[width=0.45\textwidth]{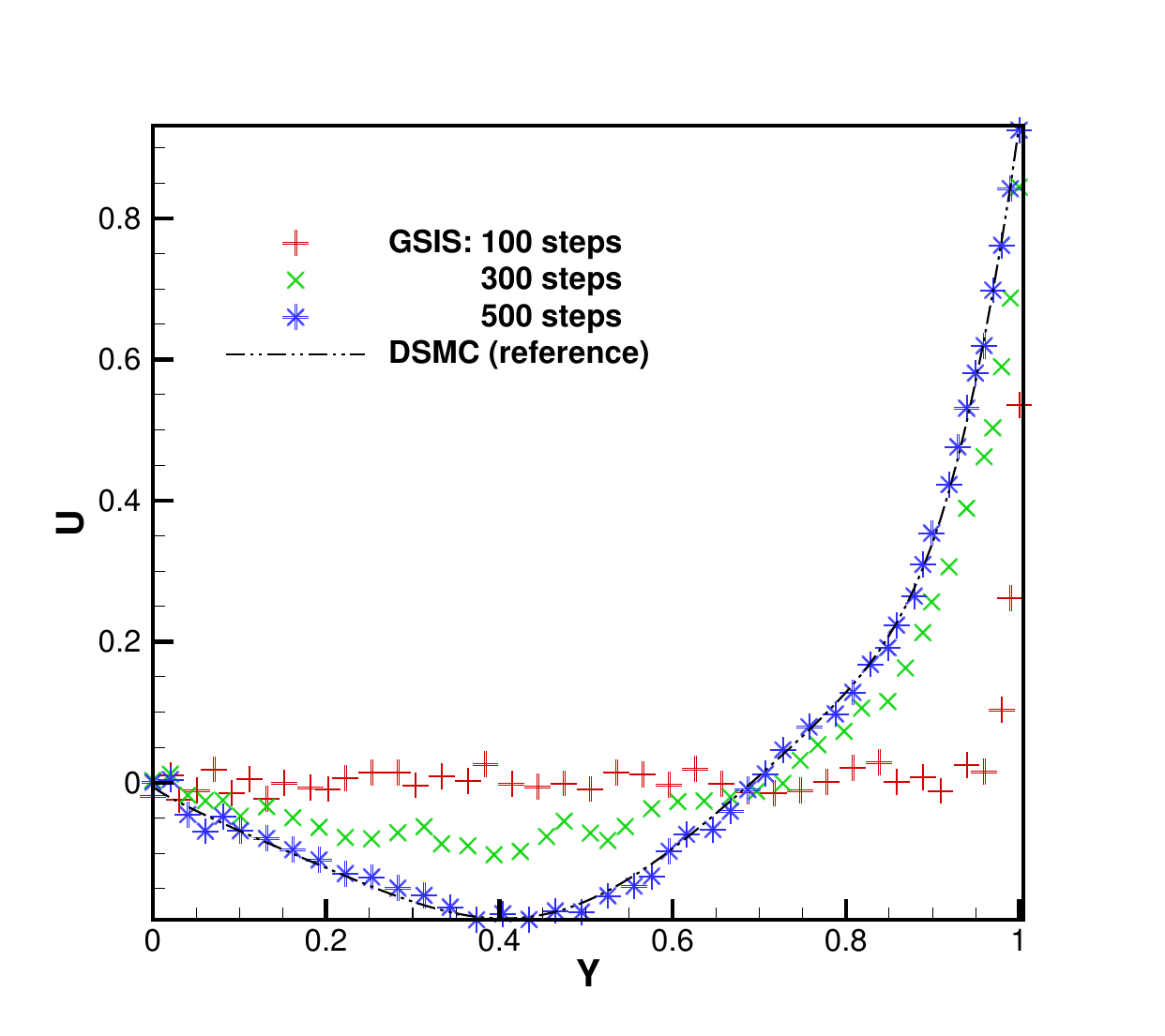}
	\caption{Comparison in the evolution of horizontal velocities between the DSMC/AAP-TRMC (left column) and GSIS (right column), when $\text{Kn}=0.1$ (top) and 0.01 (bottom).}
	\label{Cavity_evolution_figure}
\end{figure}

%


Figure~\ref{Cavity_evolution_figure} shows the evolution of the flow velocity in the x-direction along the horizontal center line. 
When $\text{Kn} = 0.1$, a consistent time step is utilized across all three schemes. The DSMC and AAP-TRMC require approximately 800 steps to reach the steady state, whereas GSIS necessitates just 300 steps.
However, when the Knudsen number decreases, since DSMC employ smaller grid size $\Delta x$ and time step $\Delta t$, GSIS exhibits a notably superior acceleration effect in comparison to DSMC.
Here, we only compare the iteration steps required for reaching the steady state in GSIS and AAP-TRMC schemes, employing identical cell size and time step.
When $\text{Kn}=0.01$, the AAP-TRMC method requires nearly 7000 iteration steps, while the GSIS needs only 500 steps, demonstrating its superior computational efficiency when the Knudsen number is relatively low.

\begin{figure}[!h]
	\centering
	\subfloat{\includegraphics[width=0.45\textwidth]{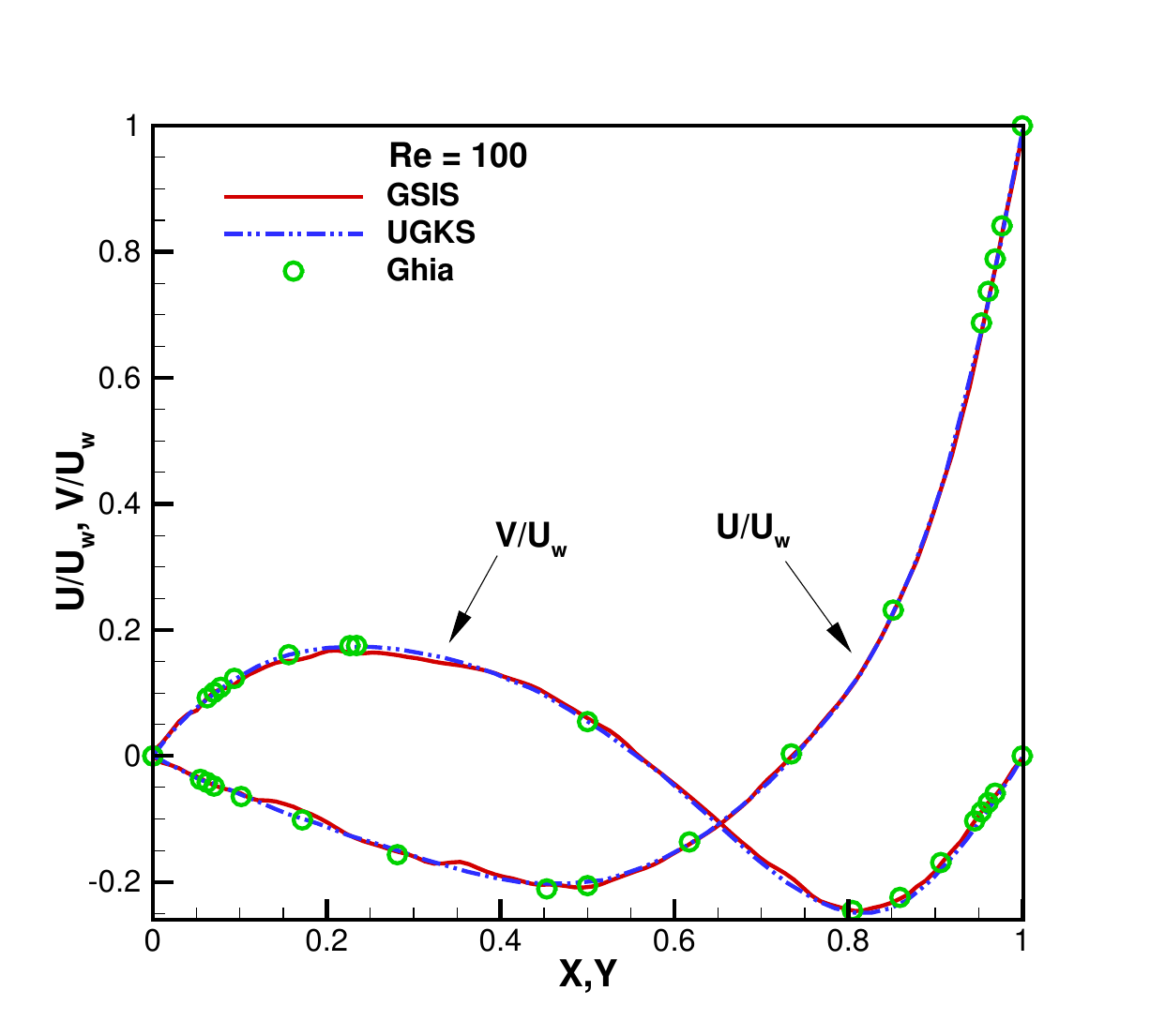}\label{Re100UoverUw_compare}}
	\subfloat{\includegraphics[width=0.45\textwidth]{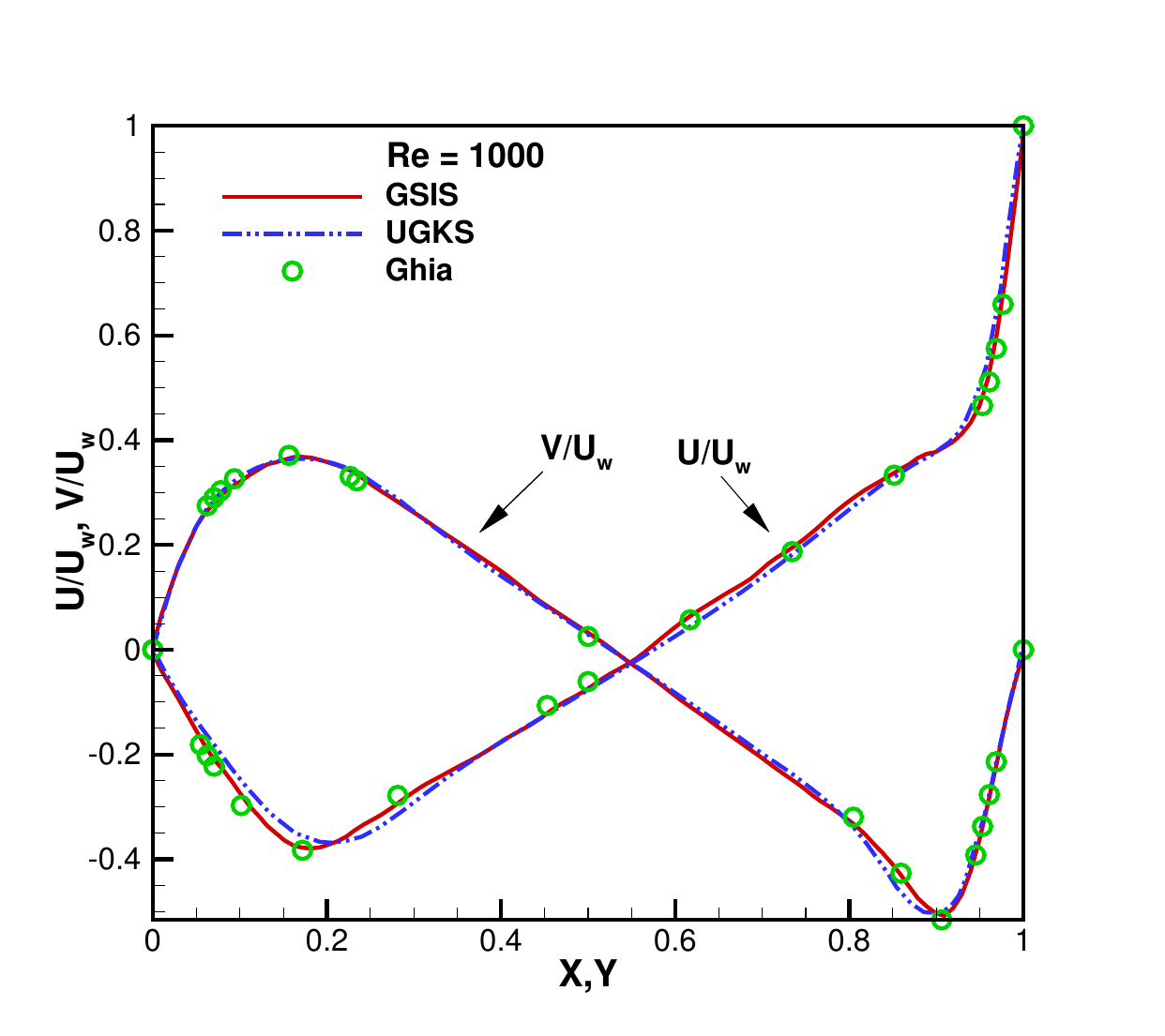}\label{Re1000UoverUw_compare}}
	\caption{The profile of the horizontal (vertical) velocity along $x=0.5$ ($y=0.5$) lines of the cavity with different Reynolds number (Left: Re = 100; Right: Re = 1000). Ghia's data are obtained from Ref.~\cite{ghia-1982} and UGKS results are extracted from Ref.~\cite{huang-2012}}.
	\label{Cavity_ReUV_figure}
\end{figure}

Figure~\ref{Cavity_ReUV_figure} compares the GSIS results with Ghia's benchmark CFD data \cite{ghia-1982} and UGKS results \cite{huang-2012}, demonstrating its capability of simulating gas flows in the near-continuum regime.
For cases with $\text{Re} = 100 $ and $\text{Re} = 1000 $, the same $150\times150$ grids with the first layer thickness $\Delta x_{\min}=0.001$ are employed.
In the near-continuum regime, the computational cost of DSMC is unaffordable. And due to the explicit time-stepping scheme, the computational cost of AAP-TRMC scheme is still expensive. However, by coupling with GSIS, the evolution of particles can reach the steady state easily.

Table~\ref{tab2} compares the number of steps employed as well as the CPU time for the three schemes. When $\Kn=0.1$ and 1, the overall CPU cost of GSIS is larger due to extra cost in solving synthetic equations. However, as the Knudsen number decreases, GSIS needs significantly fewer steps and CPU times than AAP-TRMC and DSMC. When $\Kn=0.01$, the CPU time in the transition state is reduced by 58 and 11 times when compared to DSMC and AAP-TRMC, respectively. The CPU time required for GSIS in the steady state is approximately halved compared to other two schemes. Particularly, when $\text{Re}=100$ and 1000, GSIS significantly outperforms the AAP-TRMC, with a CPU time reduction of an order of magnitude in the transition state and twofold in the steady state, demonstrating its enhanced convergence in the near-continuum regime.

\subsection{Hypersonic flows passing over a cylinder}

\begin{figure}[!t]
	\centering
	\includegraphics[width=0.45\textwidth]{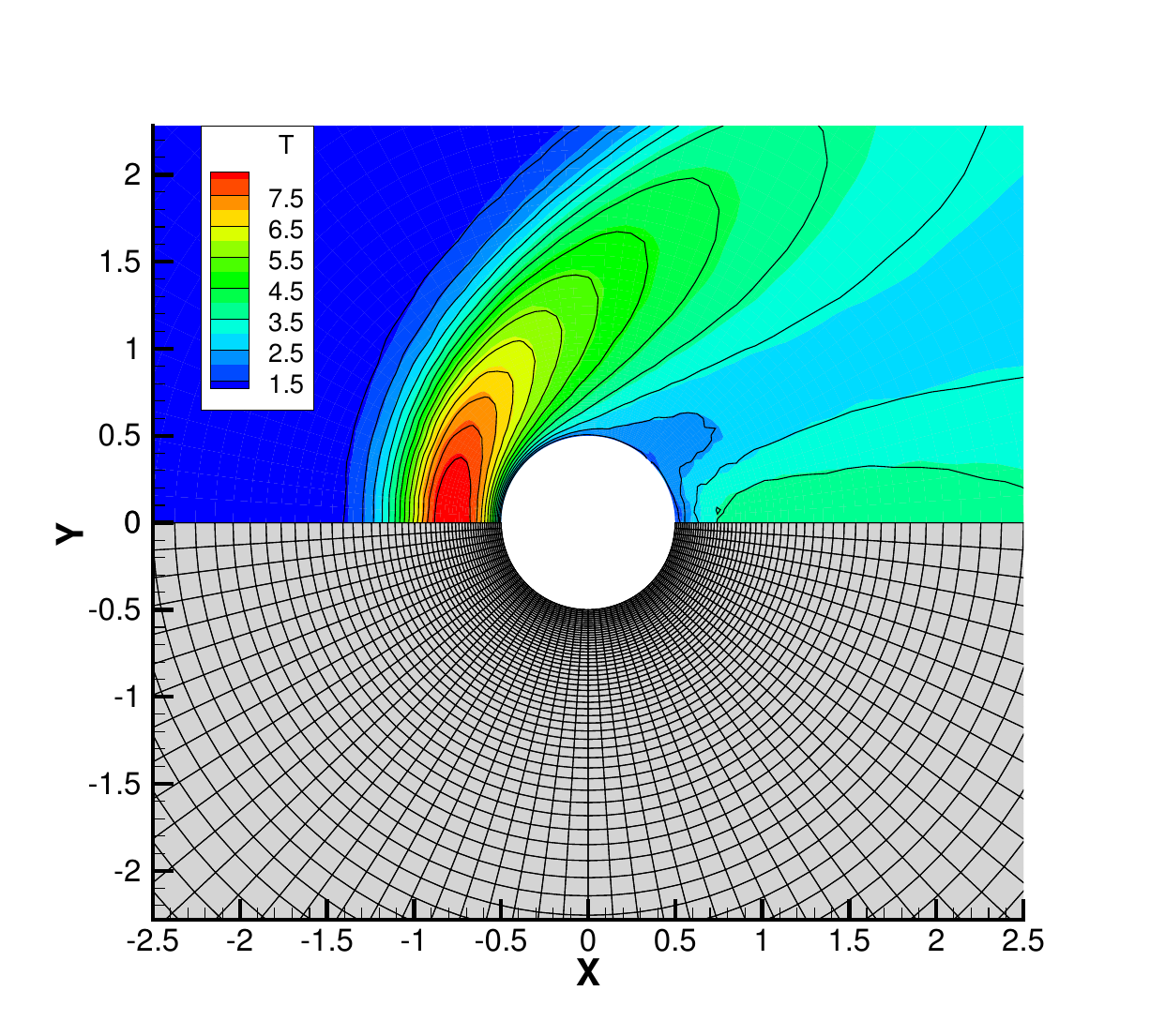}
	\includegraphics[width=0.45\textwidth]{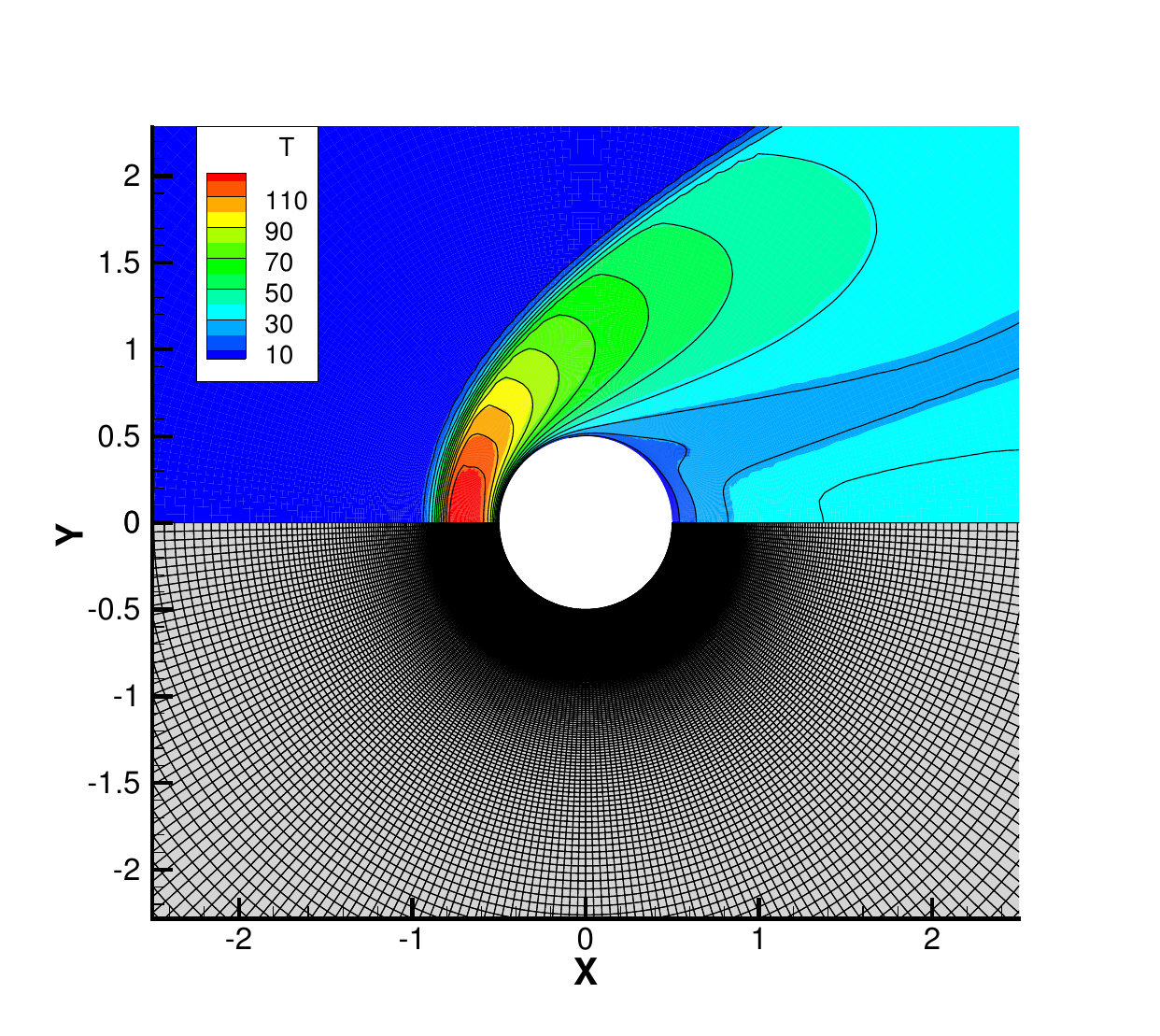}
	\caption{The mesh and the temperature contour for cylinder flow at $\text{Ma}=5, Kn=\text{0.1}$ (left) and $\text{Ma}=20, \text{Kn}=0.01$ (right). Contours: GSIS; Lines: DSMC. 
 }
	\label{2Dwithcells}
\end{figure}

The computational domain is an annulus with a inner circle being the cylinder surface (with normalized wall temperature $T_w=1$) and outer circle being the far field. The radius of the outer circle is $5.5L$, and that of the inner circle is $0.5L$. 
The Knudsen number is defined according to the cylinder diameter $L$, while other parameters such as the number density, temperature and viscosity of the free-stream are chosen as reference values. 
The total cell numbers in the circumferential and radial directions are $M$ and $N$, respectively. As shown in Fig.~\ref{2Dwithcells}, when $\text{Kn}=0.1$, the non-uniform structured grid is set as $M=100$ and $N=64$, with the length of the first layer $\Delta h = 0.2\lambda$. When $\text{Kn}=0.01$, the physical grid is set as $M=300$, $N=200$ and $\Delta h=0.2\lambda$. For all cases, 200 particles are assigned in each cell initially.
Furthermore, when $\text{Kn}=0.1$, an identical CFL number is employed in three numerical schemes, specified at 0.2, and is increased to 0.5 for GSIS and AAP-TRMC schemes when $\text{Kn}=0.01$.

\begin{table}[!t]
 \centering
 \caption{\label{tab3}Computational overhead of the DSMC, AAP-TRMC, GSIS (the upper, middle, and lower rows of each Knudsen number, respectively) for the hypersonic cylinder flow. The computational time is given in core$\cdot$hours. Simulations denoted with a superscript $*$ were performed with non-uniform Cartesian grids and uniform initial conditions by the SPARTA program, while others are initialized by solutions to NS equations with no-slip boundary conditions.}
\begin{threeparttable} 
  \begin{tabular}{c c c  c c c c c}\toprule
 \multirow{2}{*}{Kn} &  \multirow{2}{*}{Ma}  & \multirow{2}{*}{CFL} & \multirow{2}{*}{$N_{\text{cell}}$}  & \multicolumn{2}{c}{Transition state}  &  \multicolumn{2}{c}{Steady state} \\ \cmidrule(r){5-6} \cmidrule(r){7-8}
 ~ & ~ & ~ &~ & steps&time&steps   &    time\\ \hline
 \multirow{3}{*}{0.1} &  \multirow{3}{*}{5} &\multirow{3}{*}{0.2} & $100\times 64$  &    700   & 0.76 &  10000 &      11     \\
~   & ~  & ~ & $100\times 64$    &  700   & 1.32 &  10000 &      18     \\
~   & ~  & ~ & $100\times 64$   &  400   & 0.83 &  5000 &      11     \\  \addlinespace
 \multirow{3}{*}{0.01} & \multirow{3}{*}{5}  & 0.2&  2010616$^*$  &    -   & 300 &  - &  295     \\
~   & ~       & 0.5 & $300\times 200$ &    7000   & 92 &  5000 &      68    \\
~   & ~       & 0.5 & $300\times 200$ &    400   & 4.8 &  3000 &      47     \\  \addlinespace
 \multirow{3}{*}{0.01} & \multirow{3}{*}{10} & 0.2 &  2010616$^*$  &    -   & 200 &  - &  356     \\
~   & ~       &0.5& $300\times 200$ &    5000   & 72 &  3000 &      44     \\
~   & ~       &0.5& $300\times 200$ &    300   & 5.6 &  1000 &     19     \\  \addlinespace
 \multirow{3}{*}{0.01} & \multirow{3}{*}{20}  & 0.2 &2010616$^*$  &    -   & 200 &  - &  368     \\
~   & ~       &0.5& $300\times 200$ &    3000   & 51 &  3000 &      51     \\
~   & ~      &0.5& $300\times 200$ &    300   & 5.8 &  1000 &     21     \\  \addlinespace
\bottomrule
\end{tabular}
 \end{threeparttable}
\end{table}

\begin{figure}[!t]
	\centering
	\includegraphics[width=0.45\textwidth]{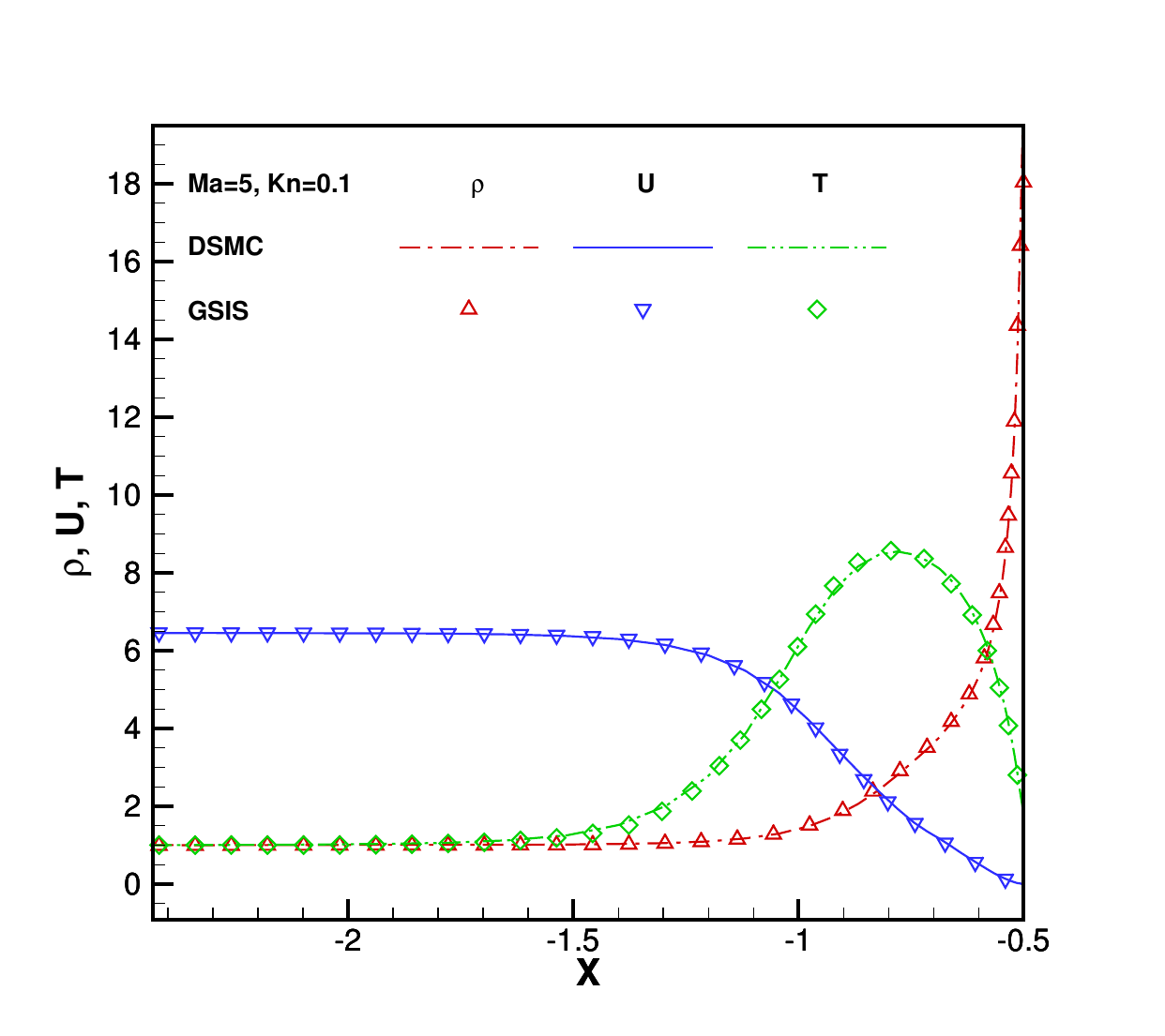}
	\includegraphics[width=0.45\textwidth]{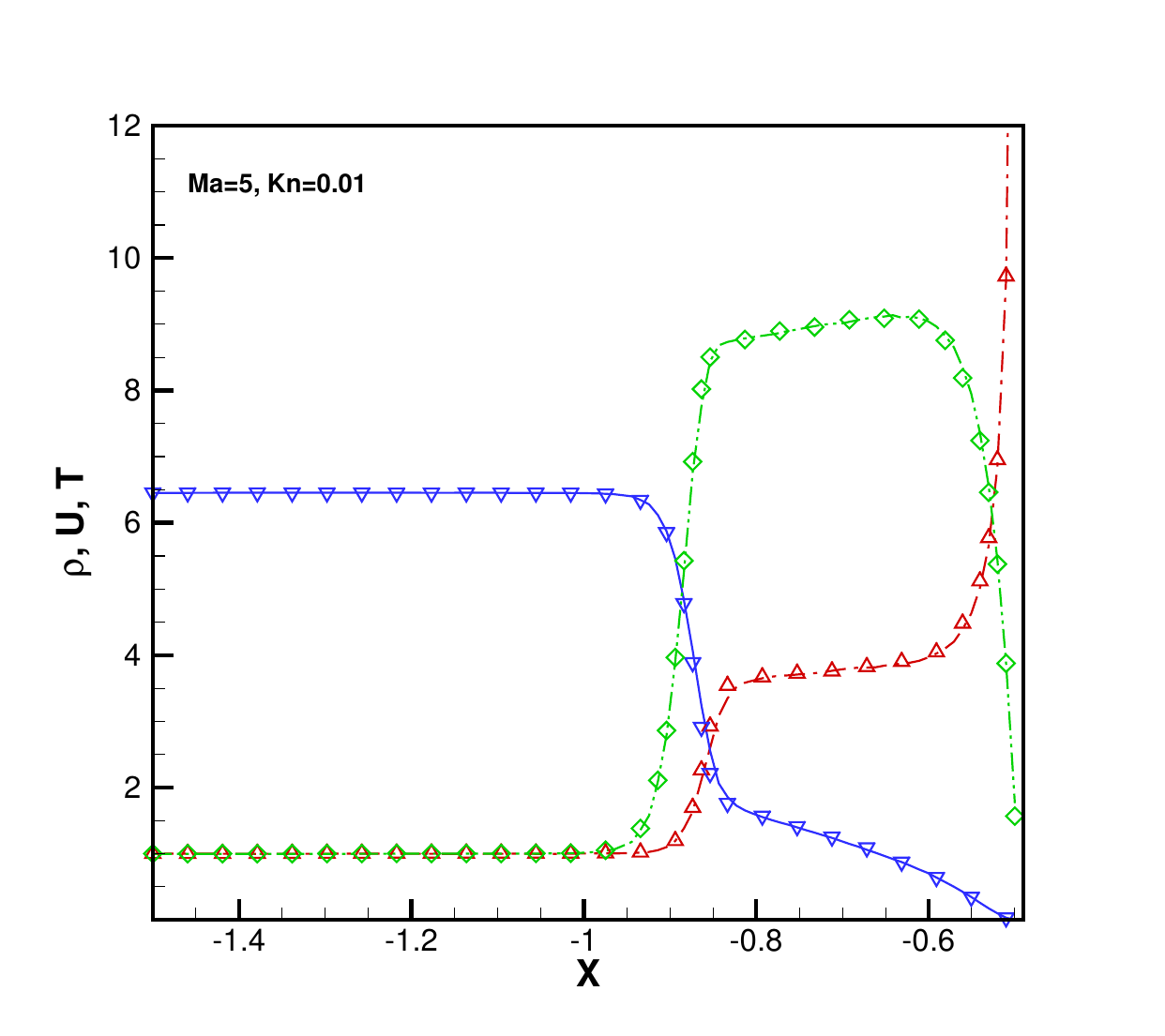}\\
	\includegraphics[width=0.45\textwidth]{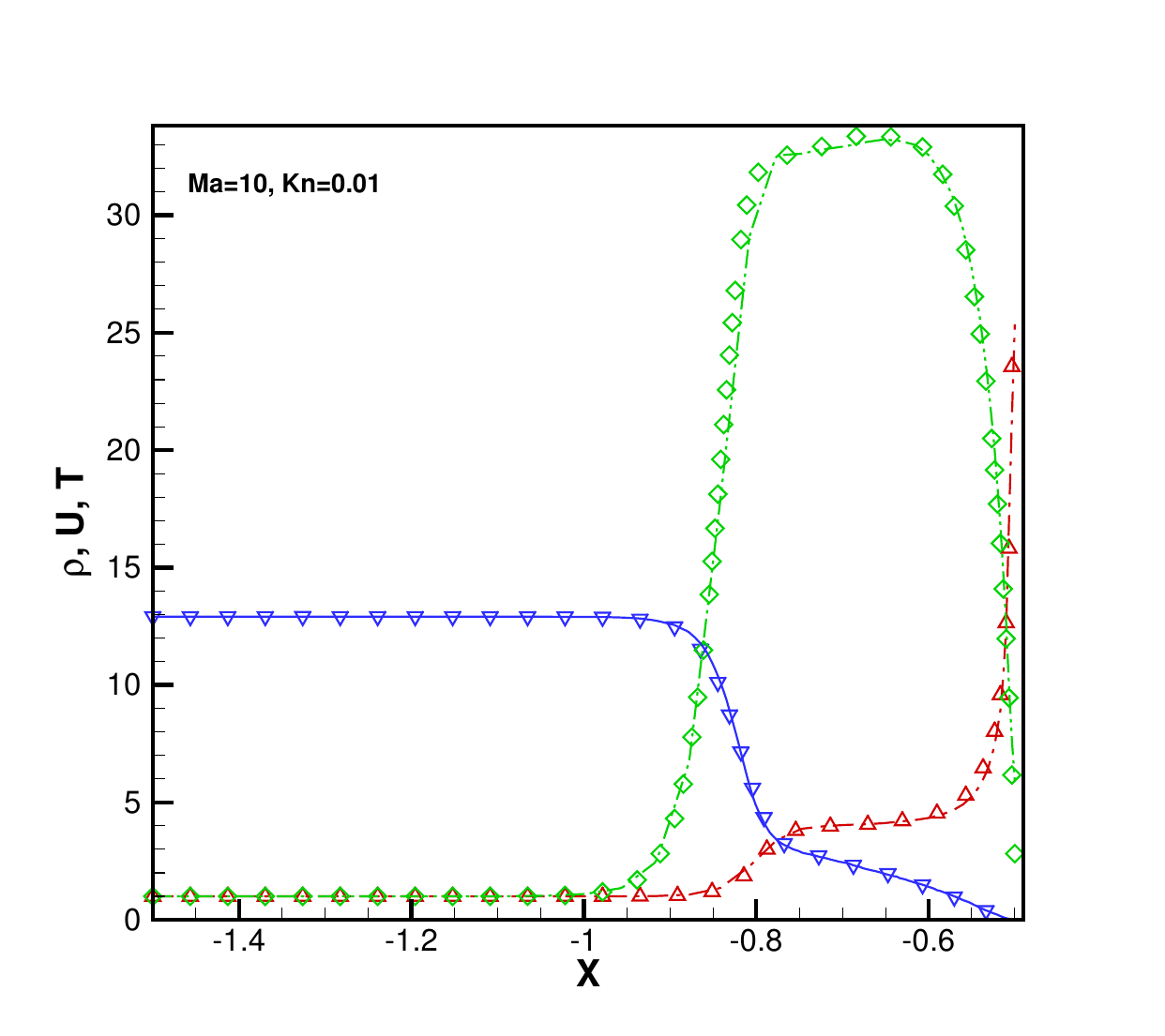}
	\includegraphics[width=0.45\textwidth]{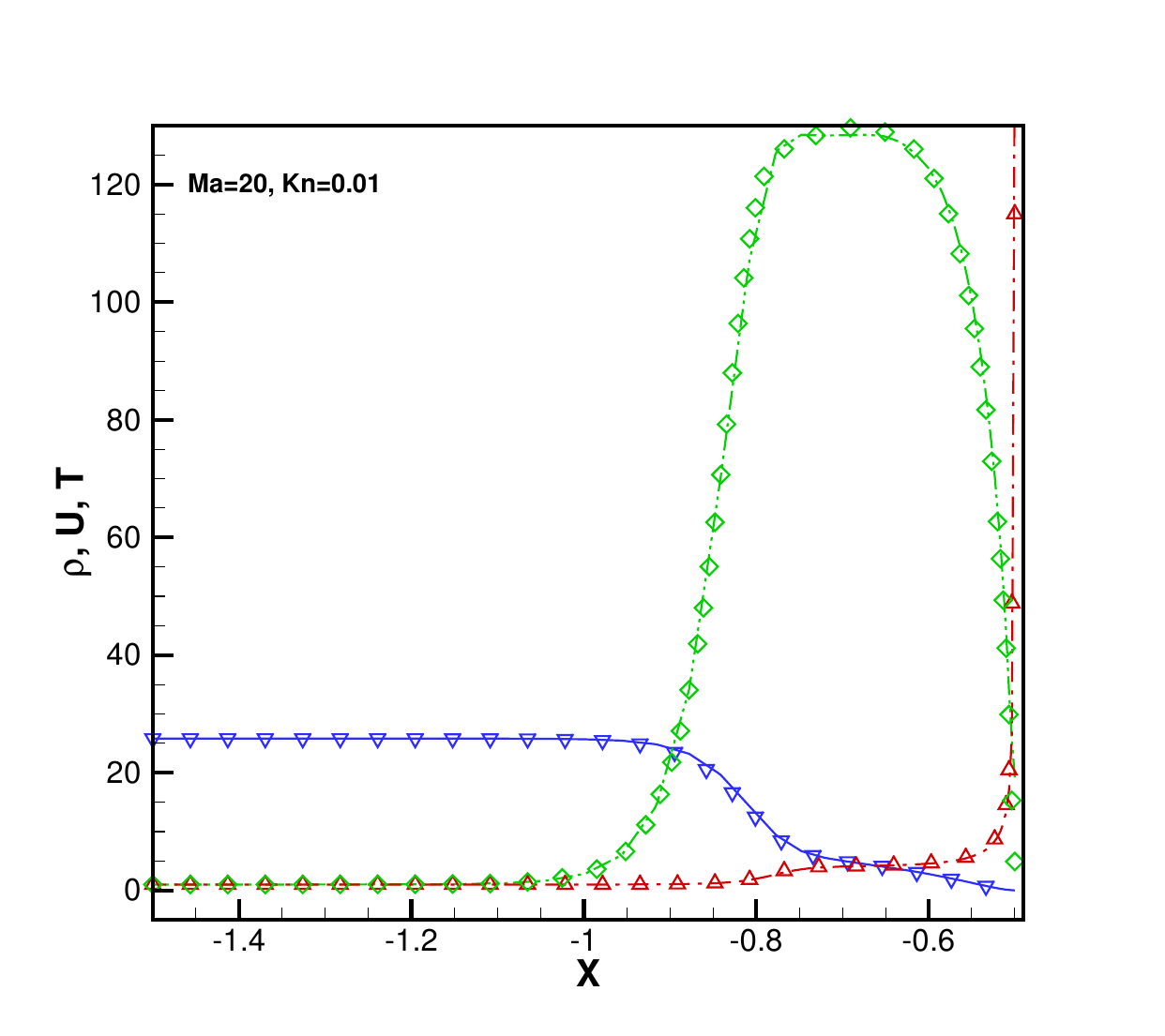}
	\caption{The distribution of macroscopic properties along the horizontal central line for the gas flow passing over a cylinder in different flow conditions.}
	\label{1DMa}
\end{figure}

Figure \ref{2Dwithcells} compares the temperature contour for different Mach numbers and Knudsen numbers, demonstrating the consistency between GSIS and DSMC results. 
Moreover, the macroscopic properties along the stagnation stream line in windward side of the cylinder are extracted and compared in Fig.~\ref{1DMa}. It is apparent that the shock wave preceding the cylinder diminishes in thickness as the Knudsen number diminishes. And as Mach number increases, the temperature ahead of the cylinder surface exhibits a significant rise, culminating in a maximum value of approximately $125T_{ref}$ for $\text{Ma}=20$.


\begin{figure}[!t]
	\centering
	\includegraphics[width=0.45\textwidth]{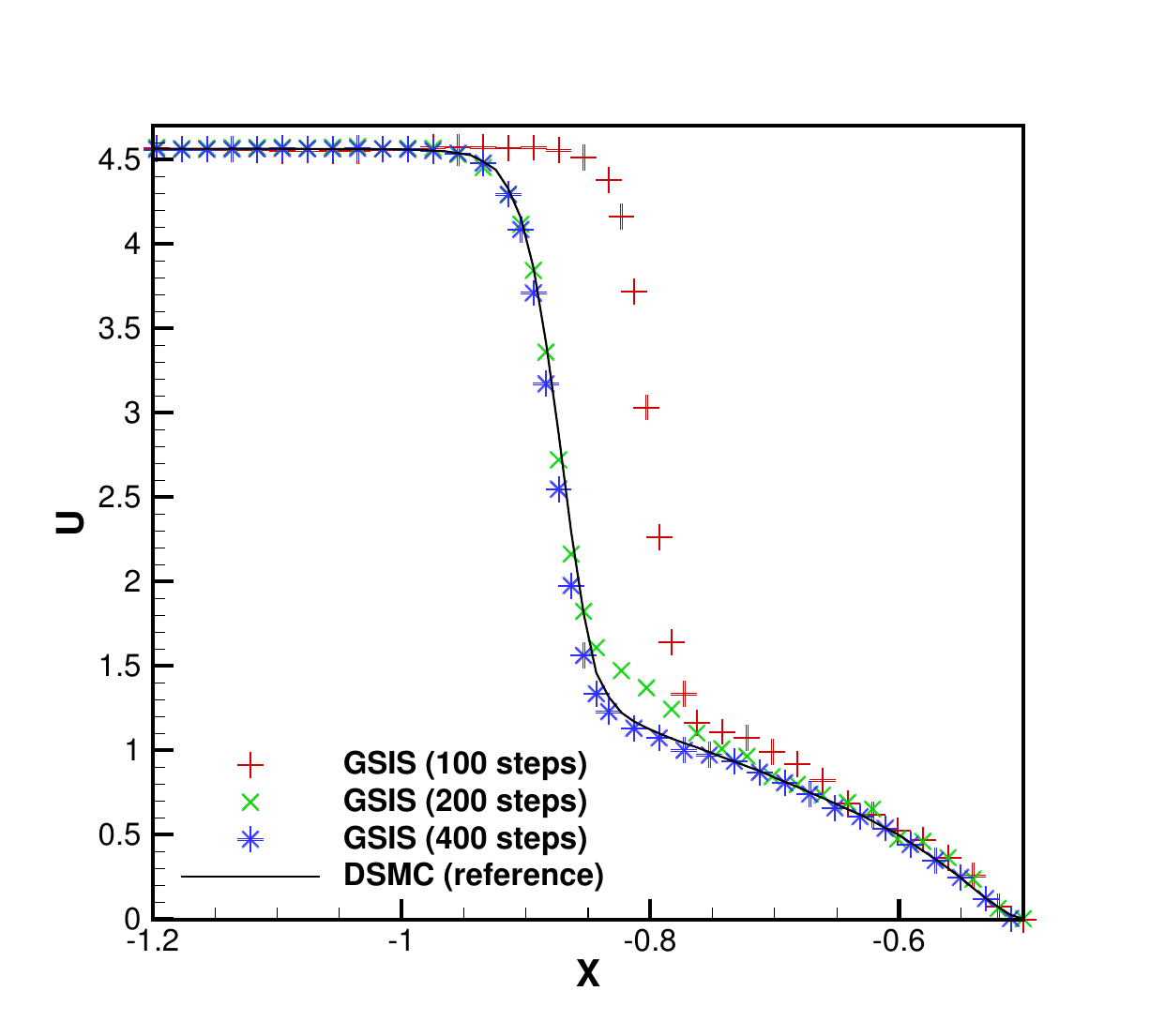}
	\includegraphics[width=0.45\textwidth]{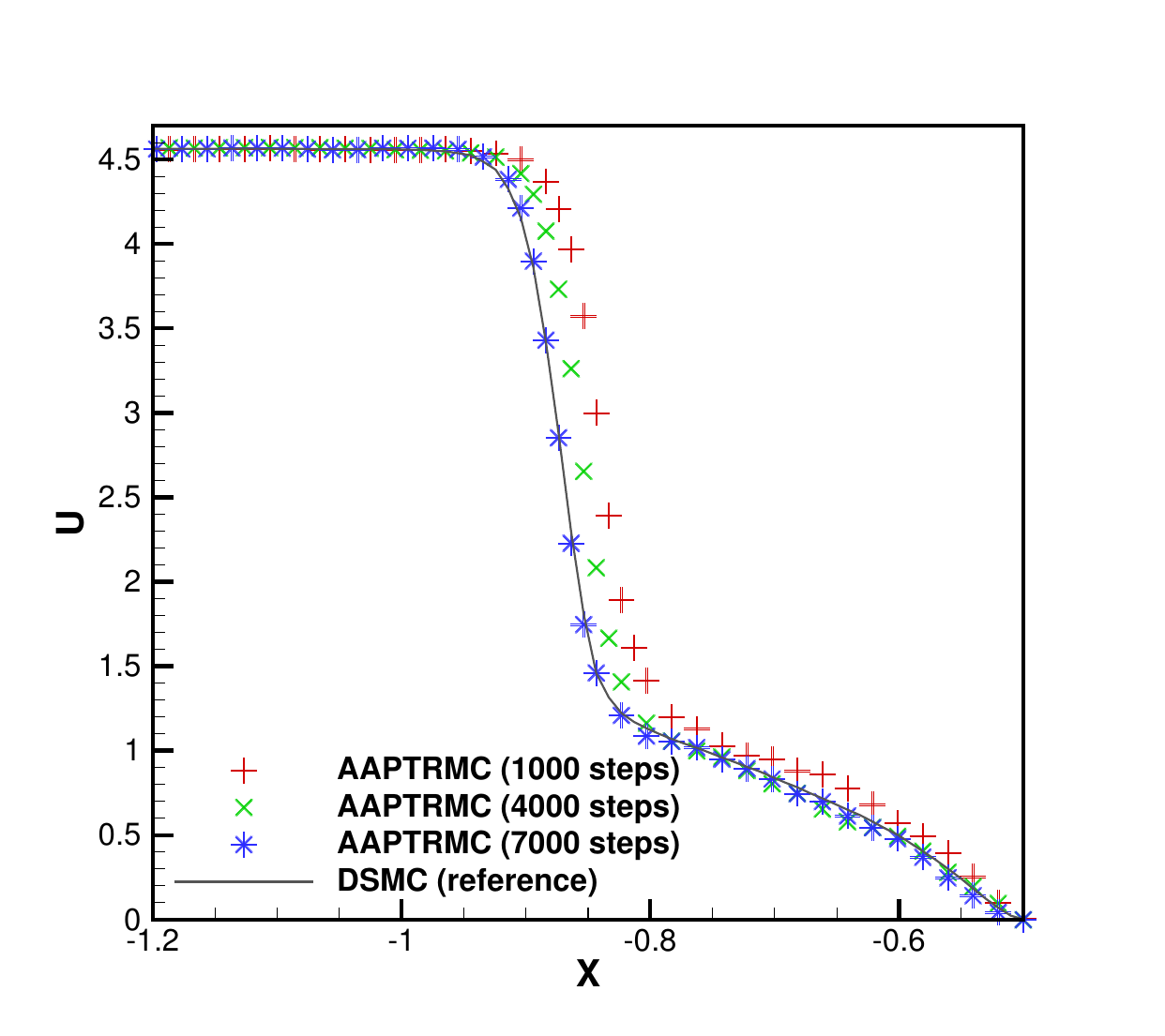}
	\caption{The evolution of the velocity in x-direction along the leading edge of the cylinder when $\text{Ma}=5,\,Kn=0.01$ for different schemes. The initial field for both two schemes is given by identical solutions to NS equations with no-slip boundary conditions.}
	\label{Kn001Ma5evolution}
\end{figure}

Figure \ref{Kn001Ma5evolution} depicts the evolution of the velocity in the x-direction along the leading edge of the cylinder when $\text{Ma}=5, \text{Kn}=0.01$. 
Due to the restrictions on the grid size and time step, the number of steps required in the transition state for DSMC is significantly larger than GSIS.
Here, we focus solely on the evolution in the transition state for GSIS and AAP-TRMC, which employ identical grid size and time step. 
As depicted in Fig.~\ref{Kn001Ma5evolution}, AAP-TRMC necessitates nearly 7000 steps to reach the steady state, whereas GSIS requires approximately 400 steps, resulting in a reduction of one order of magnitude in the number of iteration steps. Note that both schemes are initialized with identical NS results, where the no-slip boundary condition is implemented. 

The CPU time for different schemes are compared in Table~\ref{tab3}. When $\text{Ma}=5$ and $\Kn=0.1$, 
the overall CPU time for GSIS is nearly the same as that of the DSMC.
However, the fast convergence is achieved as the Knudsen number decreases. 
When $\Kn=0.01$, since the reference DSMC data are obtained by non-uniform Cartesian grids with uniform initial condition in SPARTA, only the CPU time is recorded in Table~\ref{tab3}.
For different Mach numbers, GSIS shows its fast evolution towards the steady state, with nearly one order of magnitude reduction in CPU time compared to the AAP-TRMC, and fourty times reduction compared to the DSMC. 
In the steady state, due to the continuous guidance of particle evolution, GSIS requires fewer time-averaged samples. Consequently, the CPU time is approximately halved compared to the AAP-TRMC scheme.

 \section{Conclusions and outlooks}\label{sec:6}

In summary, the GSIS has been effectively integrated with the AAP-TRMC approach to simulate the rarefied gas dynamics based on the Boltzmann equation. The method exhibits consistent accuracy across all levels of gas rarefaction, with notably fast convergence in near-continuum flow regimes. This is accomplished by iteratively solving the macroscopic synthetic equations and the mesoscopic Boltzmann equation using the AAP-TRMC technique. Following a designated sampling period in the AAP-TRMC process, the time-averaged macroscopic properties are obtained, along with the constitutive relations for stress and heat flux that include higher-order terms. These are then fed into  the macroscopic systhetic equations to get a solution which is more close to the final steady state.
The macroscopic properties are subsequently employed to refine particle velocities in AAP-TRMC through two distinct mechanisms: firstly, by adhering to the principles of mass, momentum, and energy conservation, particle velocities are adjusted via a linear transformation, steering the flow towards a steady state. Secondly, within the same sampling interval, the revised macroscopic properties influence the distribution function of the sampled particles, ensuring the preservation of the AP characteristic of the scheme.
A series of numerical experiments have confirmed that, in the near-continuum regime, the GSIS requires significantly fewer iterations to achieve a steady state, reducing the count by at least one to two orders of magnitude in comparison to the explicit time-stepping schemes of both AAP-TRMC and DSMC methods.

Theoretically, the proposed acceleration technique can be readily adapted for use with other direct simulation Monte Carlo methods with AP property (e.g.,  the exponential Runge-Kutta method \cite{dimarco-2011} and asymptotic-preserving method \cite{ren-2014}). However, it is essential that these methods are first modified to incorporate the asymptotic-preserving property for the Navier-Stokes equations to ensure the accuracy of the macroscopic solutions. It can also be used to accelarate the time evolution of the unified gas-kinetic wave-particle methods~\cite{LiuZhu-2020}, where the simulation particles gradually diminish as the Knudsen number decreases.
Furthermore, given that the DSMC models are more adept at integrating physicochemical processes, the foundational concept of this method is well-suited for accelerating the simulaiton of multi-scale multi-physics rarefied gas flows.

\section*{Acknowledgments}

This work is supported by the National Natural Science Foundation of China (12172162) and the Stable Support Plan 80000900019910072348. 


\bibliographystyle{elsarticle-num}
\bibliography{ref}

\end{document}